\documentclass[intlimits,a4paper]{article}
\pdfoutput=1
\usepackage{jheppub}
\usepackage{graphicx}
\usepackage{amsmath}
\usepackage{amssymb}
\usepackage{microtype}
\usepackage{url}
\usepackage[utf8]{inputenc} 
\usepackage[english]{babel}

\title{\boldmath Probing the small-$x$ nuclear gluon distributions with isolated photons at forward rapidities in p+Pb collisions at the LHC}

\author{Ilkka Helenius,}
\author{Kari J. Eskola and}
\author{Hannu Paukkunen} 

\affiliation{Department of Physics, University of Jyv\"askyl\"a, P.O. Box 35, FI-40014 University of Jyv\"askyl\"a, Finland}
\affiliation{Helsinki Institute of Physics, P.O. Box 64, FIN-00014 University of Helsinki, Finland}

\emailAdd{ilkka.helenius@jyu.fi}
\emailAdd{kari.eskola@phys.jyu.fi}
\emailAdd{hannu.t.paukkunen@jyu.fi}

\abstract{
Inclusive direct photon production in p+Pb collisions at the LHC is studied within the NLO perturbative QCD. Our aim is to quantify the dominant $x$ regions probed at different rapidities and to identify the best conditions for testing the nuclear gluon parton distribution functions (nPDFs) at small $x$. A comparison to the inclusive pion production reveals that from these two processes the photons carry more sensitivity to the small-$x$ partons and that this sensitivity can be further increased by imposing an isolation cut for the photon events. The details of the isolation criteria, however, seem to make only a small difference to the studied $x$ sensitivity and have practically no effect on the expected nuclear modifications. We consider also the yield asymmetry between forward and backward rapidities which can be used to probe the nPDFs irrespectively of whether an accurate p+p baseline is available.
}


\begin{document}

\maketitle
\flushbottom

\section{Introduction}

Within collinear factorization \cite{Collins:1989gx,Brock:1993sz} the inclusive cross section to produce a hard elementary particle $k$ in a collision of hadrons $h_1$ and $h_2$ can be calculated as
\begin{equation}
\mathrm{d} \sigma^{h_1 + h_2 \rightarrow k + X}(\mu^2,Q^2) = \sum\limits_{i,j,X'} f_{i/h_1}(x_1,Q^2) \otimes f_{j/h_2}(x_2,Q^2) \otimes \mathrm{d}\hat{\sigma}^{ij\rightarrow k + X'}(\mu^2,Q^2), 
\end{equation}
where the parton distribution functions (PDFs) $f_{i/h_1}(x_1,Q^2)$ $(f_{j/h_2}(x_2,Q^2))$ describe the number density distributions of partons $i$ ($j$) in a hadron $h_1$ ($h_2$) at a momentum fraction $x_1(x_2)$ and factorization scale $Q$. The piece $\mathrm{d}\hat{\sigma}^{ij\rightarrow k + X'}$ can be calculated as a perturbative expansion in strong and electroweak couplings. The dependence on the renormalization scale $\mu$ is indicated. The PDFs are non-perturbative and cannot currently be calculated from the first principles of QCD. Instead, the information on the PDFs comes mainly from experimental hard-process data through global analyses \cite{Forte:2013wc}. Here, our focus will be on the nuclear PDFs (nPDFs) and prospects of resolving the differences with respect to the free-nucleon PDFs. 

The majority of the data that are used to constrain the nPDFs at the present global fits \cite{Eskola:2009uj, Hirai:2007sx, deFlorian:2011fp, Schienbein:2009kk, Kovarik:2013sya} (see refs.~\cite{Eskola:2012rg, Paukkunen:2014nqa} for recent reviews) are from fixed-target deep inelastic scattering (DIS) and low-mass Drell-Yan dilepton measurements and have remained almost the same since the first public parametrization \cite{Eskola:1998df}. While these data offer direct constraints for the quarks, the nuclear gluons remain only weakly constrained, mostly indirectly through the DGLAP \cite{Lipatov:1974qm,Gribov:1972ri,Altarelli:1977zs,Dokshitzer:1977sg} scale evolution and the momentum sum rule. The most recent available global next-to-leading order (NLO) fits, EPS09 \cite{Eskola:2009uj} and DSSZ \cite{deFlorian:2011fp}, exploit also the RHIC data for inclusive pion production in d+Au collisions at mid-rapidity to obtain more direct gluon constraints in the region $x>0.01$. Both analyses involve also Hessian uncertainty studies \cite{Pumplin:2001ct} resulting with PDF error sets which can be used to quantify how the nPDF uncertainties propagate to physical observables and estimate the impact of new experimental measurements \cite{Paukkunen:2014zia}. Although there are significant differences among independent sets of nPDFs, we will consider here only EPS09 which appears consistent with the first p+Pb jet measurements at the LHC \cite{Chatrchyan:2014hqa} and which also has the largest uncertainties of the available parametrizations.

\begin{figure}[htb]
\centering
\includegraphics[width=\textwidth]{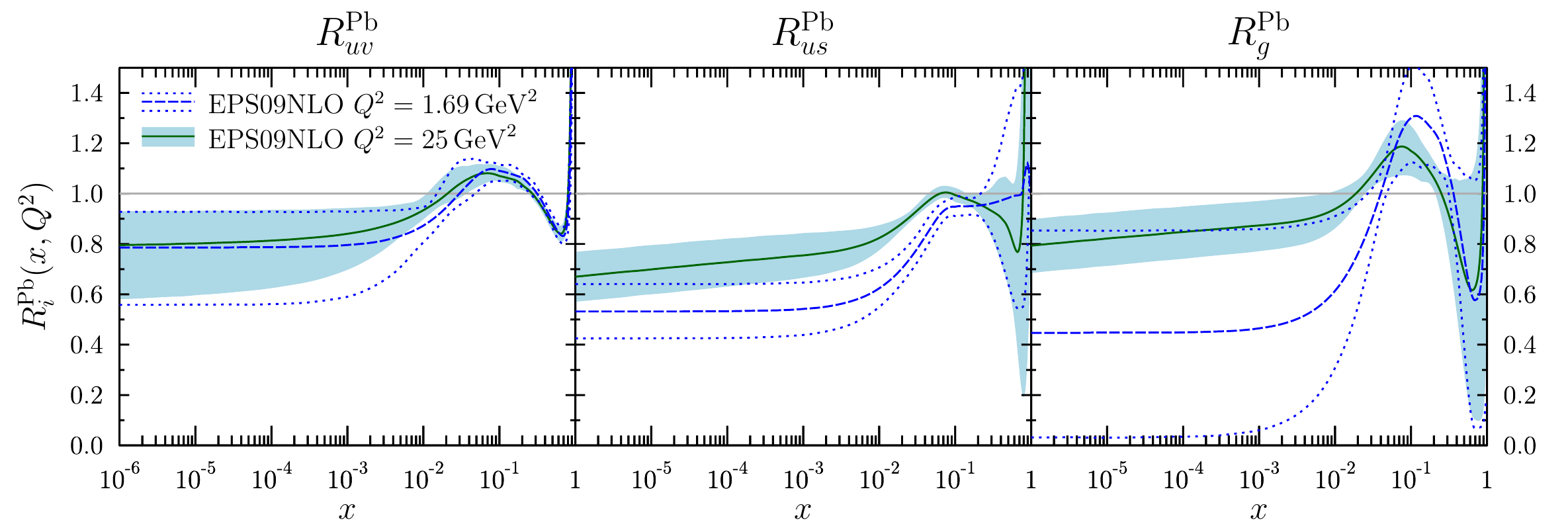}
\caption{The NLO nuclear modification for valence $u$-quarks (left), sea $u$-quarks (middle) and gluons (right) of a lead nucleus at $Q^2=25\,\mathrm{GeV^2}$, and their uncertainties, from the EPS09 analysis. The dashed curves are for the EPS09 initial scale $Q_0^2=1.69\,\mathrm{GeV^2}$, and the dotted lines show the uncertainties at $Q_0^2$.}
\label{fig:R_i_x}
\end{figure}

In figure \ref{fig:R_i_x} we show the nuclear modifications of the up valence quarks $R_{u_V}$, up sea quarks $R_{u_s}$, and gluons $R_{g}$, at $Q^2=1.69$~GeV$^2$ and (relevant for our discussion below) $Q^2=25$~GeV$^2$  for lead nucleus as predicted by EPS09. The nuclear quarks appear rather well constrained wherever they dominate the measured DIS and DY processes, i.e. at $x\gtrsim 0.1$ for valence quarks and at $0.01\lesssim x \lesssim 0.1$ for sea quarks. However, it should be borne in mind that these modifications were assumed to be flavor independent at the parametrization scale $Q^2=1.69$~GeV$^2$ and involve a rather restricted functional form below $x\sim 10^{-2}$, which leads to an underestimation of the true uncertainty. Although the nuclear gluons have much less data constraints the DGLAP evolution is observed to quickly shrink the originally extensive error bands at $x\lesssim 0.1$. On one hand, this property makes the DGLAP-based predictios rather robust in the sense that there cannot be a strong suppression in observables sensitive to small-$x$ gluons at large $Q^2$. On the other hand, to further constrain the small-$x$ nuclear gluons, very precise measurements will be needed, which may be difficult to obtain from other than the clean DIS environment \cite{Paukkunen:2013hbm}.

In the near future, the most promising source for new nPDF constraints are the hard processes in p+Pb collisions at the LHC \cite{Kang:2012am,Brandt:2014vva,Guzey:2012jp,Paukkunen:2010qg,Eskola:2013aya,Arleo:2004gn,Arleo:2011gc,Stavreva:2010mw,Dai:2013xca}.
With the naive leading order (LO) $2 \rightarrow 2$ kinematics one can estimate the nuclear-side $x$ (that is, $x_2$) from
\begin{equation}
x_2=\frac{q_T}{\sqrt{s_{NN}}}\left[\mathrm{e}^{-\eta_1} + \mathrm{e}^{-\eta_2}\right] \stackrel{\eta_1\approx\eta_2\approx\eta}{\approx} \frac{2q_T}{\sqrt{s_{NN}}}\mathrm{e}^{-\eta},
\label{eq:x_approx}
\end{equation}
where $q_T$ is the transverse momentum of the produced partons and $\eta_1,\eta_2$ their rapidities. Thus to probe small $x_2$ one should consider collisions with large center-of-mass energy $\sqrt{s_{NN}}$ and/or observables at large $\eta$. In this work our goal is to quantify in detail the $x_2$ regions probed by inclusive direct photon production at different rapidities and transverese momenta $p_T$, according to the NLO calculations with LHC kinematics. In addition, we study the effect of an isolation cut and briefly discuss the inclusive hadron production for comparison. The direct photons at forward rapidities as a probe of gluon nPDFs were proposed earlier in ref.~\cite{Arleo:2007js}. Here we also extend this LO study to NLO level, accounting for the nPDF uncertainties which are nowadays available. Related studies on the direct photon production in nuclear collisions at the LHC have appeared earlier \cite{Arleo:2004gn, Arleo:2011gc, Albacete:2013ei, Stavreva:2010mw, Dai:2013xca}, also in the context of centrality dependence \cite{Helenius:2013bya}. Some aspects presented here have relevance also for the PDF studies in p+p collisions \cite{d'Enterria:2012yj} as well as for the search for the onset of non-linear effects \cite{Mueller:1985wy,Eskola:2002yc} and parton saturation
\cite{Gribov:1984tu} built into the color-glass-condensate (CGC) framework \cite{McLerran:1993ni} (see refs.~\cite{Albacete:2014fwa, JalilianMarian:2012bd, Rezaeian:2012ye, Albacete:2013ei}).
Further motivation for the present study is provided by a proposal to install a forward calorimeter (FoCal) to the ALICE detector which could measure the isolated photons with an accuracy better than  $10\,\%$ at the $3<\eta<5$ region and $p_T\ge5\,\mathrm{GeV/c}$ \cite{T.PeitzmannfortheALICEFoCal:2013fja}. To coincide with these ALICE plans, we perform the calculations here at the nominal center-of-mass energy of the LHC p+Pb collisions, $\sqrt{s_{NN}}=8.8\,\mathrm{TeV}$. The rapidity shift due to the asymmetric collision system is not considered, all our results quoted below are in the nucleon-nucleon center-of-mass system.

\section{Inclusive hadron production}

\begin{figure}[htb]
\begin{minipage}[t]{0.48\linewidth}
\centering
\includegraphics[width=\textwidth]{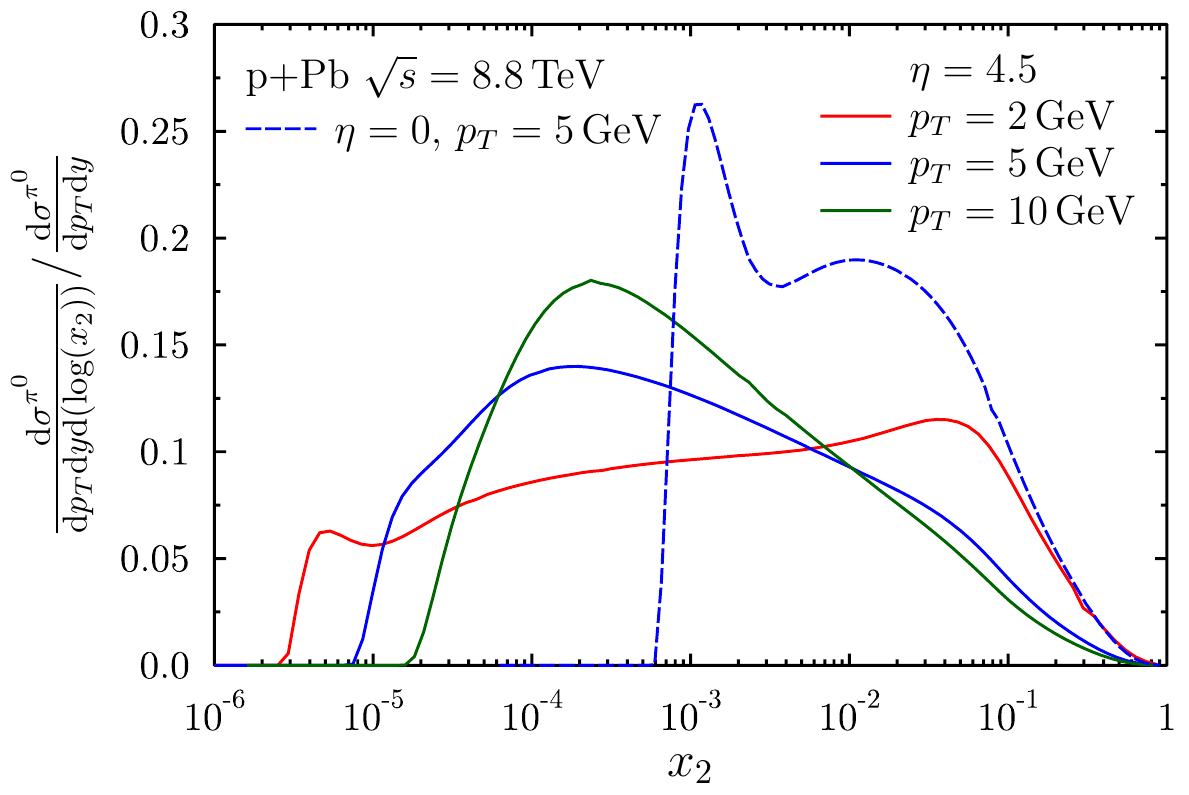}
\caption{The $x_2$ distribution for $\pi^0$ production in p+Pb collisions at $\sqrt{s_{NN}}=8.8\,\mathrm{TeV}$ and $\eta=0$ for $p_T=5\,\mathrm{GeV/c}$ (blue dashed), and at $\eta=4.5$ for $p_T=2\,\mathrm{GeV/c}$ (red), $p_T=5\,\mathrm{GeV/c}$ (blue) and $p_T=10\,\mathrm{GeV/c}$ (green).}
\label{fig:dsigma_x2_pion}
\end{minipage}
\hspace{0.02\linewidth}
\begin{minipage}[t]{0.48\linewidth}
\centering
\includegraphics[width=\textwidth]{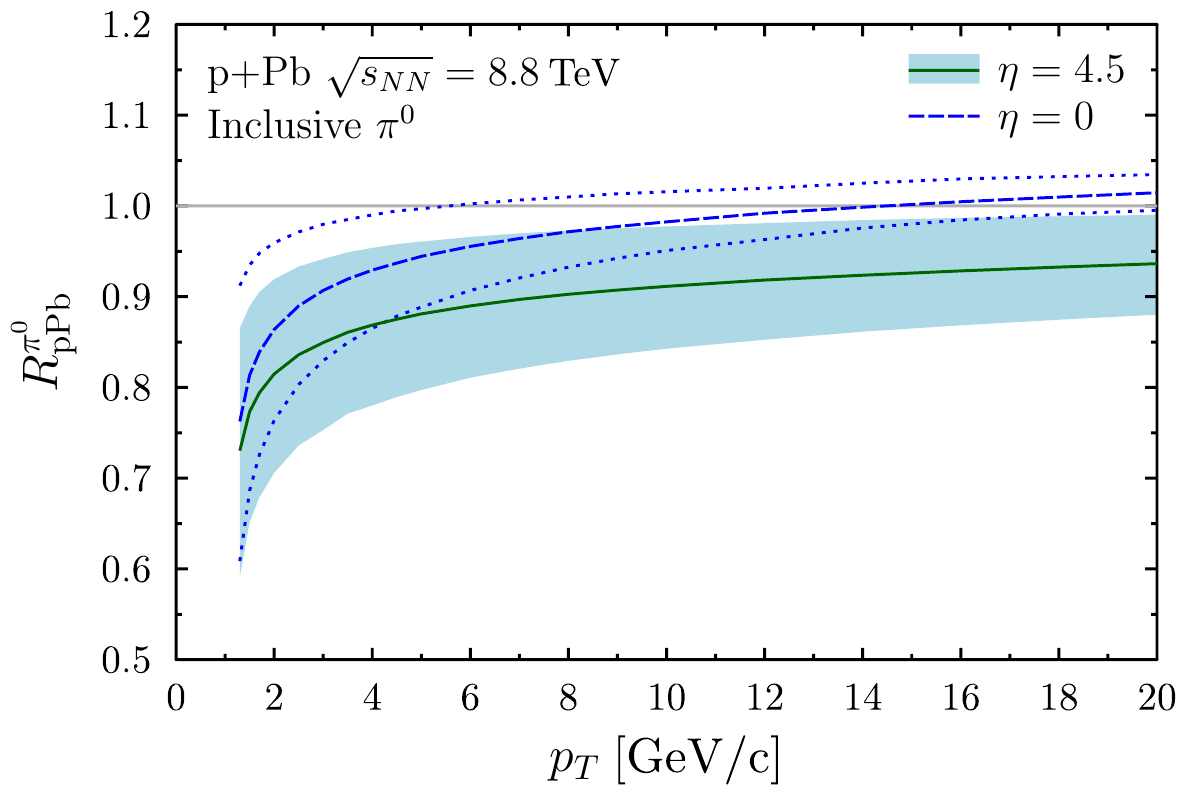}
\caption{The nuclear modification ratio $R_{\rm pPb}^{\pi^0}$ for $\pi^0$ production at $\eta=0$ (blue dashed) and  $\eta=4.5$ (green solid) using the EPS09 NLO nPDFs. The lightblue uncertainty band (the blue dotted lines for $\eta=0$) is calculated from the error sets of EPS09.}
\label{fig:R_pPb_pi0_y45}
\end{minipage}
\end{figure}

The cross section for inclusive high-$p_T$ hadron production is, loosely speaking, obtained as a convolution of the hard parton spectra and the non-perturbative parton-to-hadron fragmentation functions (FFs) $D_{h/k}(z,Q_F^2)$:
\begin{equation}
\mathrm{d} \sigma^{h+X}_{h_1h_2}(\mu^2,Q^2,Q_F^2) = \sum_k \mathrm{d} \sigma^{k+X}_{h_1h_2}(\mu^2,Q^2,Q_F^2) \otimes D_{h/k}(z,Q_F^2),
\label{eq:dsigma_h}
\end{equation}
where $z$ describes the momentum fraction carried away by the hadron $h$ from the parent parton $k$. The convolution over $z$ smears the relation between the measured final state hadron momenta $p_T$ and the partonic momenta $q_T$. Furthermore, inclusive cross sections like $\mathrm{d} \sigma/dp_Td\eta$ studied here involve integrations over the momentum fractions $x_1$ and $x_2$ such that it is not possible to access any specific value of $x_2$ but always some distribution. This is demonstrated in figure \ref{fig:dsigma_x2_pion} where we plot examples of $x_2$-distributions for differential $\pi^0$ production cross sections in p+Pb collisions at $\sqrt{s_{NN}}=8.8\,\mathrm{TeV}$ for different values of $p_T$ and $\eta$. Note that the shown cross sections are differential in $\log x_2$ (i.e. $\mathrm{d}\sigma/\mathrm{d}\mathrm{log}\,x_2=x_2\mathrm{d}\sigma/\mathrm{d}x_2$) so that the contribution from a specific $x_2$ interval can be directly read off from the $\log$-scale  in $x_2$. The NLO calculations are performed using the \texttt{INCNLO}-code \cite{incnlopage, Aversa:1988vb, Aurenche:1987fs, Aurenche:1998gv, Aurenche:1999nz} which we have modified to improve the convergence of the integrals at large $\sqrt{s_{NN}}$, large $\eta$, and small $p_T$ region\footnote{With this, we solved the numerical convergence problem which prevented one from getting reliable results in the region $p_T<10$~GeV/c at $\eta=3$ at this cms-energy e.g. at ref.~\cite{Arleo:2011gc}.}. The FFs have been taken from the DSS fit \cite{deFlorian:2007aj}, the free nucleon PDFs from CTEQ6.6M \cite{Nadolsky:2008zw} and the nuclear modifications are from EPS09 \cite{Eskola:2009uj}. The renormalization $(\mu)$, factorization $(Q)$ and fragmentation $(Q_F)$ scales are fixed to the hadron $p_T$. The uncertainties in the free proton PDFs (which are of the order $10\,\%$ for the gluons in the employed PDF set) are not considered here, since they efficiently cancel out in the nuclear cross-section ratios of our interest below.

From figure 2 one easily finds that the simple parton-level relation of eq.~(\ref{eq:x_approx}) actually corresponds rather well to the kinematic lower limit of the $x_2$ distributions, but that this or a naive estimate $\langle x_2 \rangle \approx \frac{2q_T/\langle z\rangle}{\sqrt{s_{NN}}}\mathrm{e}^{-\eta}$ with $\langle z\rangle\approx 0.5$  for the average $z$ \cite{Eskola:2002kv,Sassot:2010bh,d'Enterria:2013vba}, have no especially large contribution upon integrating over $x_2$. In fact, the cross sections $\mathrm{d} \sigma/dp_Td\eta$ get important contributions from a broad range of $x_2$.\footnote{For a similar discussion at RHIC energies, see ref.~\cite{Guzey:2004zp}.}
The peculiar shape of the $\eta=0$ result is due to the combination of the kinematical smearing in the NLO and the differentiation with respect to $\log\,x_2$ instead of $x_2$.
At forward rapidities the distributions evidently shift towards smaller values of $x_2$, as expected, but what is more surprising is that going down to very low transverse momentum, $p_T=2\,\mathrm{GeV/c}$, the relative sensitivity to smallest $x_2$ actually decreases when comparing with somewhat larger values of $p_T$.  This suggests that in searching for small-$x$ probes, instead of smallest $p_T$ one can rather focus on the region $p_T\gtrsim 5\,\mathrm{GeV/c}$, where also the pQCD framework is more reliable.

To quantify how the nuclear effects in the PDFs are expected to modify the differential cross sections and how the nPDF uncertainties propagate into these observables, we define the minimum bias nuclear modification ratio for p+Pb collisions as
\begin{equation}
R_{\rm pPb} \equiv R_{\rm pPb}(p_T,\eta) \equiv \frac{1}{208}\frac{\mathrm{d}^2\sigma_{\rm pPb}}{\mathrm{d}p_T \mathrm{d}\eta}\Big/\frac{\mathrm{d}^2\sigma_{\rm pp}}{\mathrm{d}p_T \mathrm{d}\eta},
\end{equation}
and plot it in the case of inclusive $\pi^0$ production in figure~\ref{fig:R_pPb_pi0_y45} for pseudorapidities $\eta=0$ and $\eta=4.5$ as a function of $p_T$. At $\eta=0$ we find some suppression at $p_T\lesssim 10 \,\mathrm{GeV/c}$ as the cross section is mostly sensitive to the region $x\lesssim 0.01$ which corresponds to shadowing in the EPS09 nPDFs. However, the nuclear effects are rather modest except for the very low $p_T$. Due to the smaller values of $x_2$ probed at $\eta=4.5$ we notice suppression due to the shadowing in the whole $p_T$ range considered. The nPDF-originating uncertainties at forward rapidities are larger than at $\eta=0$, which follows from the lack of direct constraints for the gluon nPDFs at $x\lesssim0.01$. The strong $p_T$ dependence of $R_{\rm pPb}$ at $p_T < 4\,\mathrm{GeV/c}$ is caused by the rapid DGLAP evolution of $R_g$ at small $Q^2$ and $x$, as was illustrated in figure~\ref{fig:R_i_x}.

\section{Direct photon production}

\begin{figure}[htb]
\begin{minipage}[t]{0.48\linewidth}
\centering
\includegraphics[width=\textwidth]{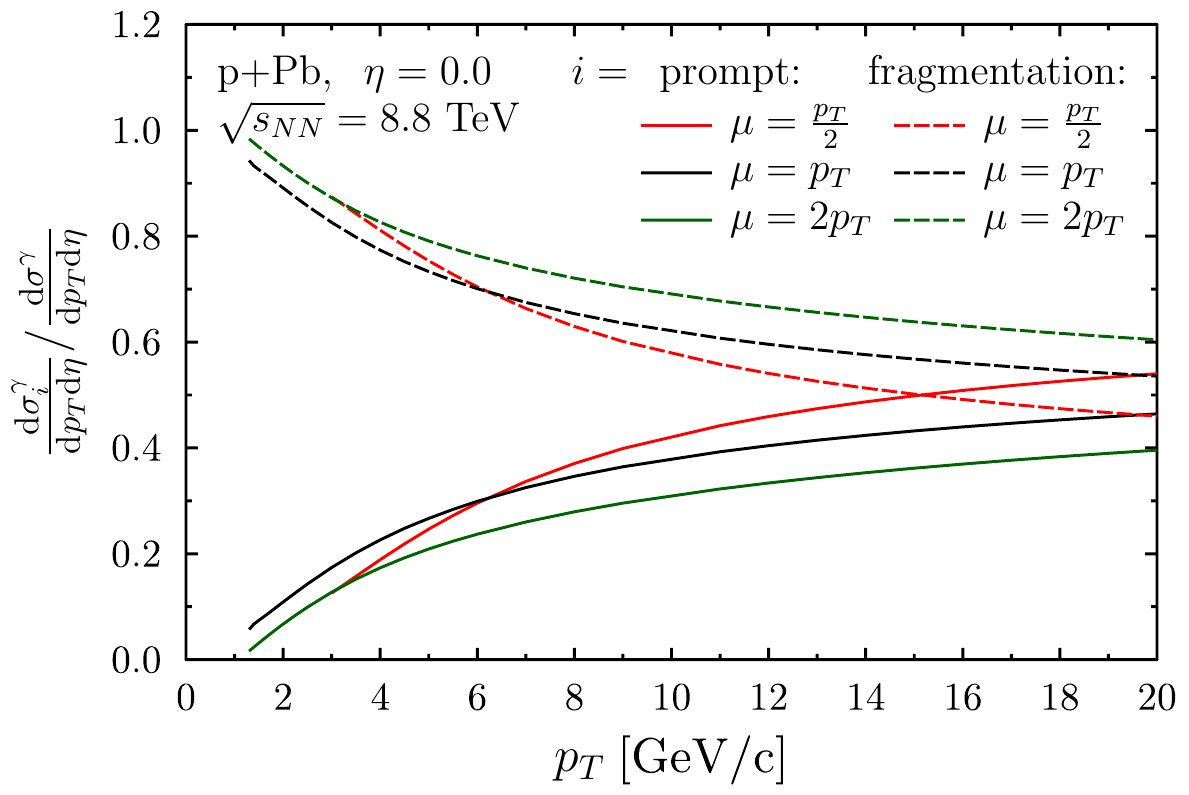}
\caption{Relative contributions of the prompt (solid) and fragmentation (dashed) components in inclusive direct photon production cross section $\mathrm{d} \sigma^{\gamma+X}_{\rm pPb}/dp_Td\eta$
as a function of the photon $p_T$ in p+Pb collisions at the LHC at $\eta=0$, with the scales fixed to $\mu=Q=Q_F=p_T/2$ (red), $p_T$ (black) and $2p_T$ (green).
}
\label{fig:frag_dir_0}
\end{minipage}
\hspace{0.02\linewidth}
\begin{minipage}[t]{0.48\linewidth}
\centering
\includegraphics[width=\textwidth]{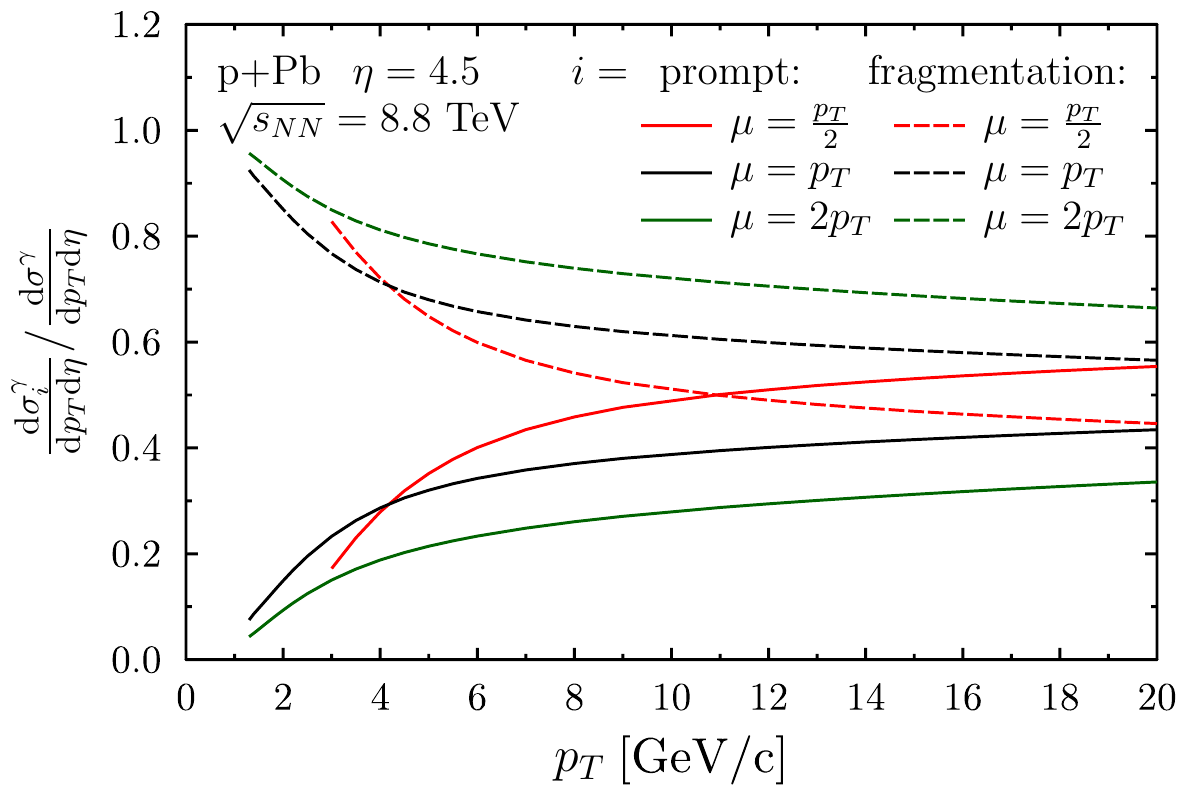}
\caption{Same as fig.~\ref{fig:frag_dir_0} but for $\eta=4.5$.}
\label{fig:frag_dir_4.5}
\end{minipage}
\end{figure}

To increase the direct small-$x_2$ sensitivity a process with a more direct access to the partonic kinematics is required. A candidate for such an observable is the prompt photon production which originates from the primary hard partonic scatterings such as the QCD Compton process. However, the experimentally measured direct photons inevitably include also the photons formed through fragmentation of the produced hard partons. Strictly speaking also in the NLO calculations the division of the direct photon production into these two components is not unambiguous but depends on the choices for the scales $\mu^2,Q^2,Q_F^2$. Thus, to compute the cross sections for what we here refer to as inclusive direct photon production, we must include contributions from both of the production mechanisms:
\begin{equation}
\mathrm{d} \sigma^{\gamma+X}_{\rm pPb} = \mathrm{d} \sigma^{\mathrm{prompt}\, \gamma+X}_{\rm pPb} + \mathrm{d} \sigma^{\mathrm{fragmentation}\, \gamma+X}_{\rm pPb},
\label{eq:dsigma_gamma}
\end{equation}
where the fragmentation component is calculated similarly to the hadron case in eq.~(\ref{eq:dsigma_h}):
\begin{equation}
\mathrm{d} \sigma^{\mathrm{fragmentation}\, \gamma+X}_{\rm pPb}(\mu^2,Q^2,Q_F^2) = \sum_k \mathrm{d} \sigma^{k+X}_{\rm pPb}(\mu^2,Q^2,Q_F^2) \otimes D_{\gamma/k}(z,Q_F^2),
\end{equation}
where $D_{\gamma/k}(z,Q_F^2)$ is now the parton-to-photon FF. Figures \ref{fig:frag_dir_0} and \ref{fig:frag_dir_4.5} show the relative contributions from these two components for the cross section $\mathrm{d} \sigma^{\gamma+X}_{\rm pPb}/dp_Td\eta$ at mid- and forward rapidity in p+Pb collisions at the LHC, with the scales fixed to $\mu=Q=Q_F=p_T/2$, $p_T$ and $2p_T$. As can be appreciated from these figures (and also noted e.g. in \cite{d'Enterria:2012yj,Chatterjee:2013naa,Klasen:2013mga} for $\eta=0$), the fragmentation photons clearly dominate at small $p_T$ in both cases and all these scale choices. The prompt component gains importance towards higher $p_T$ but the point where it becomes dominant depends on the rapidity and scale choices.

\begin{figure}[htb]
\begin{minipage}[t]{0.48\linewidth}
\centering
\includegraphics[width=\textwidth]{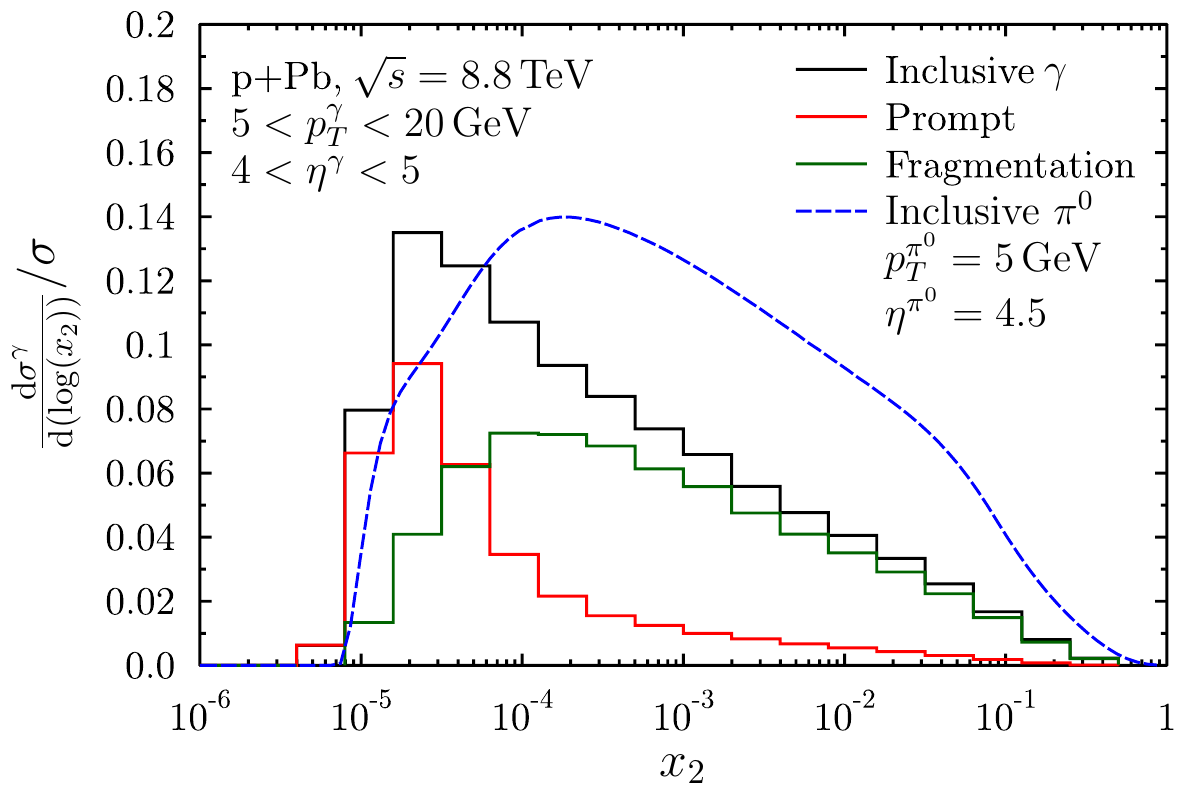}
\caption{Normalized $x_2$ distribution of inclusive $\gamma$ production (black) decomposed into the prompt (red) and fragmentation (green) components. The $\pi^0$ result from figure \ref{fig:dsigma_x2_pion} is also plotted for comparison (blue dashed). All scales are fixed to $p_T$.}
\label{fig:dsigma_x2_inc_gamma}
\end{minipage}
\hspace{0.02\linewidth}
\begin{minipage}[t]{0.48\linewidth}
\centering
\includegraphics[width=\textwidth]{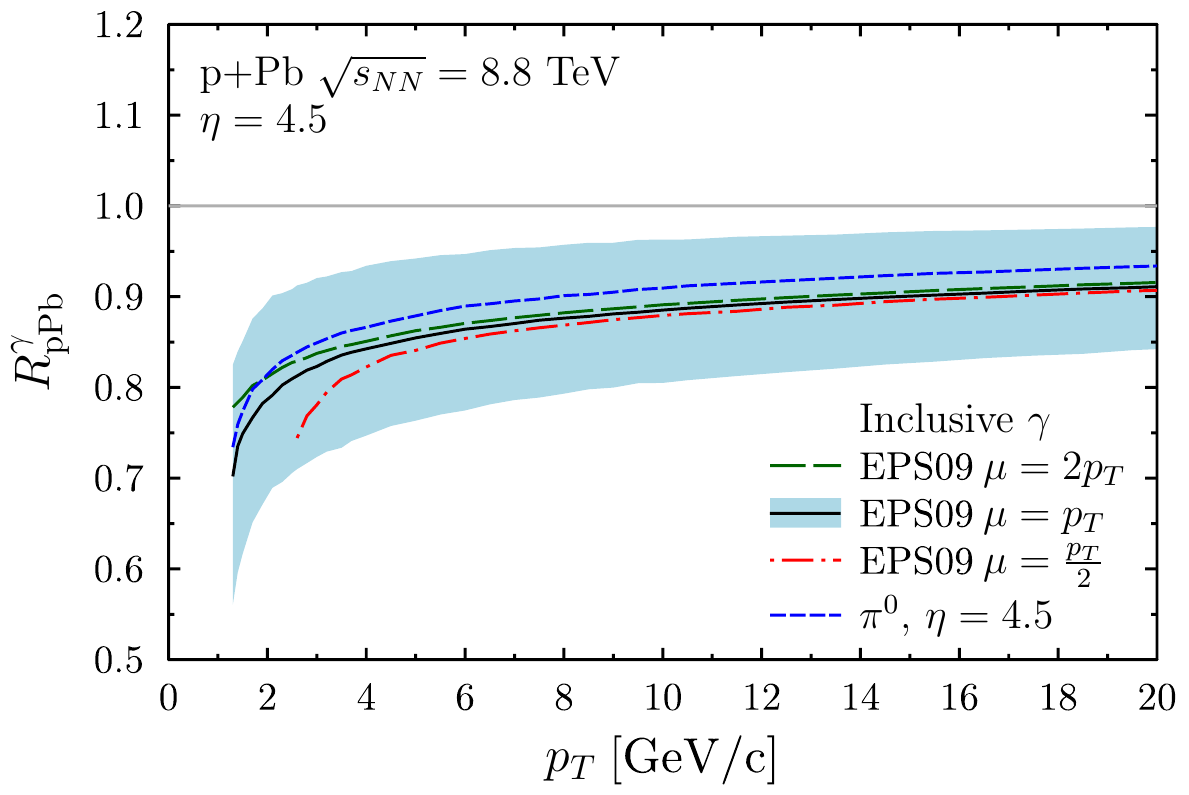}
\caption{The nuclear modification ratio $R_{\rm pPb}^{\gamma}$ for inclusive $\gamma$ production at $\eta=4.5$ with scale choices $\mu=Q=Q_F=2p_T$ (green long-dashed), $p_T$ (black solid), and $p_T/2$ (red dot-dashed) using the EPS09 nPDFs. The lightblue uncertainty band is for $\mu=Q=Q_F=p_T$. The $R_{\rm pPb}^{\pi^0}$ from figure \ref{fig:R_pPb_pi0_y45} is plotted for comparison (blue dashed).}
\label{fig:R_pPb_inc_gamma}
\end{minipage}
\end{figure}

To study the $x_2$-sensitivity of these two components we plot, in figure  \ref{fig:dsigma_x2_inc_gamma}, the 
normalized differential cross sections as a function of $x_2$ for both contributions separately. We perform the NLO calculations here for p+Pb collisions at $\sqrt{s_{NN}}=8.8\,\mathrm{TeV}$, $4 < \eta < 5$ and $5 < p_T < 20 \,\mathrm{GeV/c}$, utilizing the \texttt{JETPHOX}-program \cite{jetphoxpage, Catani:2002ny, Aurenche:2006vj} with the BFGII \cite{Bourhis:1997yu} parton-to-photon FFs, and the CTEQ6.6 PDFs with the EPS09 nuclear modifications. All scales have been chosen to coincide with the photon $p_T$.
For comparison, also the $\pi^0$ result at $p_T=5$~GeV/c, $\eta=4.5$ from figure \ref{fig:dsigma_x2_pion} is included. Clearly, the relative $x_2$ sensitivity (the shape) of the fragmentation component is very similar to that in $\pi^0$ production, but the presence  of the prompt photon component  drags the total distribution towards smaller $x_2$. The increased small-$x_2$ sensitivity has, as we demonstrate in figure \ref{fig:R_pPb_inc_gamma}, only a small impact on the nuclear modification ratio $R_{\rm pPb}^{\gamma}$ in comparison to the $\pi^0$'s: The photon suppression is only slightly stronger, which is due to the rather moderate $x$ dependence in the EPS09 nPDFs at small $x$ which, as noted earlier, tends to be a general consequence of the DGLAP dynamics. Thus, also the EPS09 error bands in the pion and photon cases are very similar. In figure \ref{fig:R_pPb_inc_gamma} we also show the effect of different scale choices, $\mu=Q=Q_F=2p_T$, and $p_T/2$. Although the scale uncertainties can be rather large in the absolute cross sections, in a ratio like $R_{\rm pPb}^{\gamma}$ these cancel out rather efficiently especially at $p_T\gtrsim 4$~GeV/c. 

\begin{figure}[htb]
\begin{minipage}[t]{0.48\linewidth}
\centering
\includegraphics[width=\textwidth]{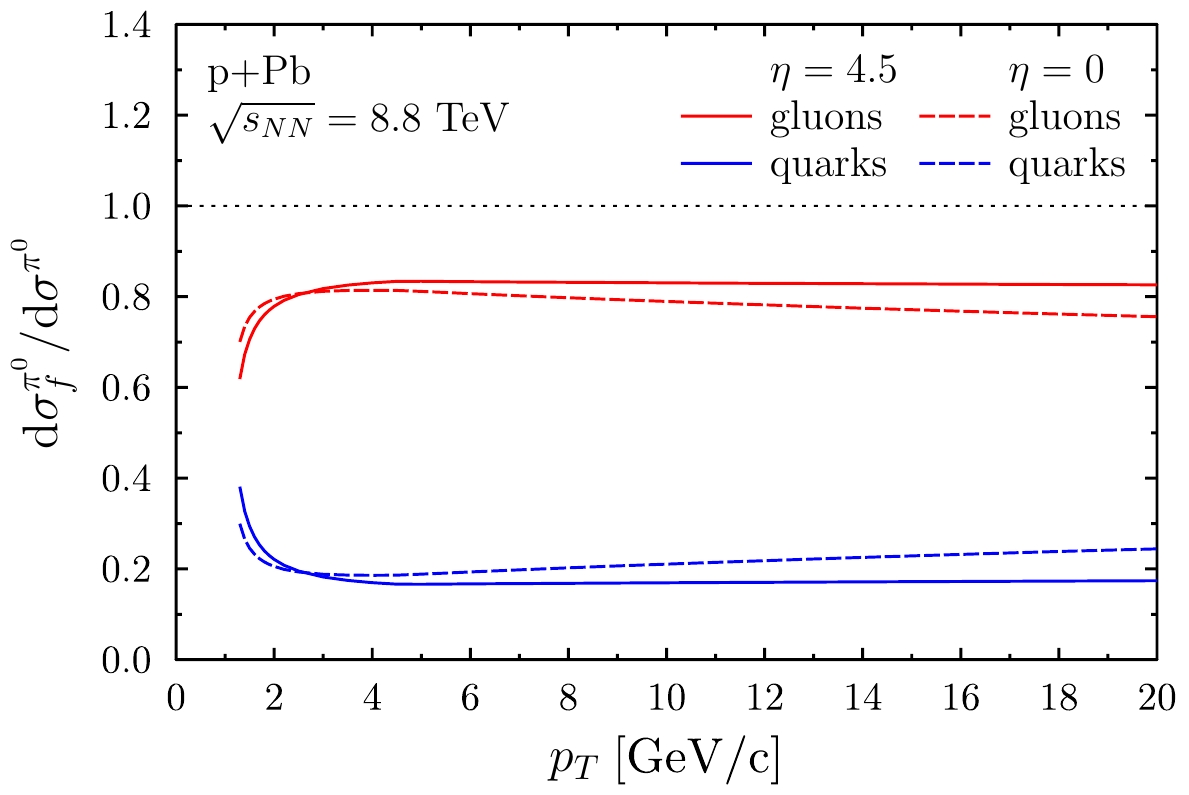}
\caption{The relative contributions from nuclear gluons (red) and quarks (blue) to inclusive $\pi^0$ cross section $\mathrm{d} \sigma^{\pi^0+X}_{\rm pPb}/dp_Td\eta$
at $\eta=4.5$ (solid) and $\eta=0$ (dashed) as a function of $p_T$.}
\label{fig:dsigma_g_vs_q_pi0}
\end{minipage}
\hspace{0.02\linewidth}
\begin{minipage}[t]{0.48\linewidth}
\centering
\includegraphics[width=\textwidth]{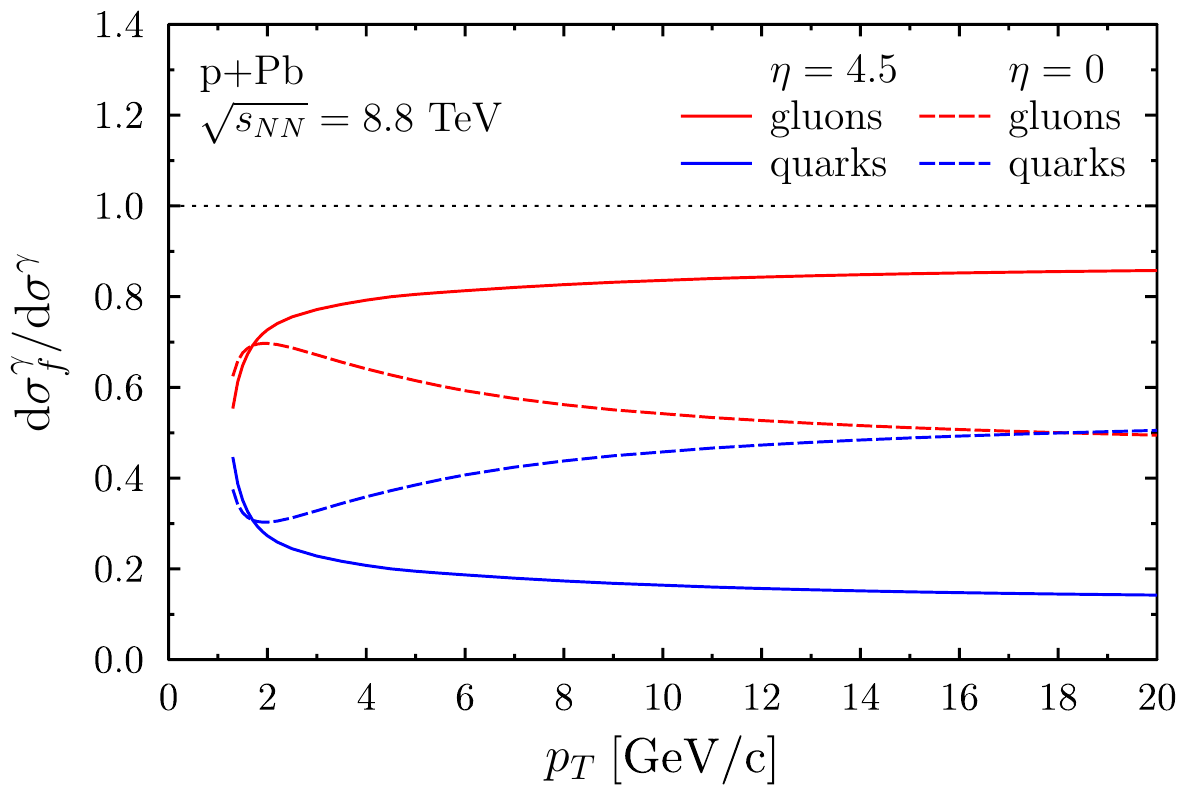}
\caption{As figure \ref{fig:dsigma_g_vs_q_pi0} but for inclusive direct photons.}
\label{fig:dsigma_g_vs_q_gamma}
\end{minipage}
\end{figure}

To check which nuclear partons are the most ``active'' ones in the inclusive particle production, the relative contributions from nuclear gluon- and quark-originating processes are shown for $\pi^0$'s in figure \ref{fig:dsigma_g_vs_q_pi0} and for direct photons in figure \ref{fig:dsigma_g_vs_q_gamma} for $\eta=0$ and $\eta=4.5$. Technically, these are obtained by setting the nuclear quark+antiquark PDFs and the gluon PDFs to zero in turn. For $\pi^0$'s the nuclear gluons generate about 80~\% of the cross sections both at mid- and forward rapidities. This is expected as the gluon PDFs dominate at $x\lesssim 0.01$ and as the gluon and quark FFs to pions are of the same magnitude. For photons the picture is different: at mid-rapidity, the nuclear quarks and gluons generate about an equal amount of the cross section but at forward rapidity the gluons again contribute at about an 80~\% level. These effects can be understood as follows:
\begin{itemize}
\item {$p_T\gtrsim 10$~GeV/c:}
The prompt photons dominate at large $p_T$, and are typically produced via Compton-like scattering $q_i+g\rightarrow \gamma + q_i$ \cite{d'Enterria:2012yj}. At $\eta=0$ the $x_2$- and $x_1$-distributions are almost identical (the nuclear effects in the nPDFs being moderate) which in practice makes it equally likely to pick a quark from either the proton or from the nucleus. At $\eta=4.5$, however, the cross sections become sensitive to smaller values of $x_2$ and larger $x_1$ so that it is more likely to pick a gluon from the nucleus and a (valence) quark from the proton. 
\item {$p_T\lesssim 10$~GeV/c:}
Unlike for hadrons, the parton-to-photon FFs are about a magnitude larger for quarks than for gluons \cite{Bourhis:1997yu}. As the fragmentation component starts to dominate in this $p_T$ region, (cf. figures \ref{fig:frag_dir_0} and \ref{fig:frag_dir_4.5}) this enhances the relative importance of the quark-initiated processes thereby partly compensating for the increasing gluon density $g(x_2)$ towards low $p_T$. For this reason the contributions from the quark and gluon initiated processes at midrapidity remain very similar also at low $p_T$.
\end{itemize}
The strong growth of the gluon contribution towards higher $p_T$ at $p_T < 2\,\mathrm{GeV/c}$ is common for pions and photons and follows from the rapid scale evolution of the small-$x$ gluon distributions close to the PDF initial scale $Q_0=1.3\,\mathrm{GeV}$. The conclusion from figure \ref{fig:dsigma_g_vs_q_gamma} is that to probe the gluon PDFs with direct photons, it is advantageous to look at the forward rapidity and $p_T\gtrsim 4$~GeV/c. 

\subsection{Isolation cut}

Although the fragmentation and prompt components cannot be measured separately in the experiments, introducing an isolation cut for the photons the fragmentation component can be suppressed. The isolation cut discards the direct photon events that have ``too much'' hadronic activity around the photon and is used in the measurements mainly to reject the background from hadronic decays, se e.g.~Refs.~\cite{Chatrchyan:2012vq, Adare:2012yt}. As the fragmentation photons are emitted collinearly to the parent parton, the isolation cut reduces the fragmentation component, making the observable more sensitive to prompt photon production and thus decreasing the probed values of $x_2$.

The most commonly used isolation criterion is to reject photon events for which the total amount of hadronic transverse energy $\Sigma E_T$ inside a cone of a fixed radius $R$, calculated as
\begin{equation}
\Sigma E_T=\sum_i E_T^i\theta(R-R_i), \text{ where } R_i=\sqrt{(\eta_{\gamma}-\eta_i)^2+(\phi_{\gamma}-\phi_i)^2},
\end{equation}
is larger than a chosen maximum $E_T^{\rm max}$. Above, $E_T^i$ is the transverse energy of the hadron $i$, $\eta_i$ ($\eta_{\gamma}$) the pseudorapidity of the hadron (photon), $\phi_i$ ($\phi_{\gamma}$) the azimuthal angle of the hadron (photon) and the sum runs over all hadrons in the event. The maximum value of the allowed $\Sigma E_T$ can be either a fixed number or it can be defined to be proportional to the photon transverse momentum. There are also other isolation criteria proposed, e.g. in ref.~\cite{Frixione:1998jh}, but here we will consider only these two types of isolation cuts.

\begin{figure}[htb]
\begin{minipage}[t]{0.48\linewidth}
\centering
\includegraphics[width=\textwidth]{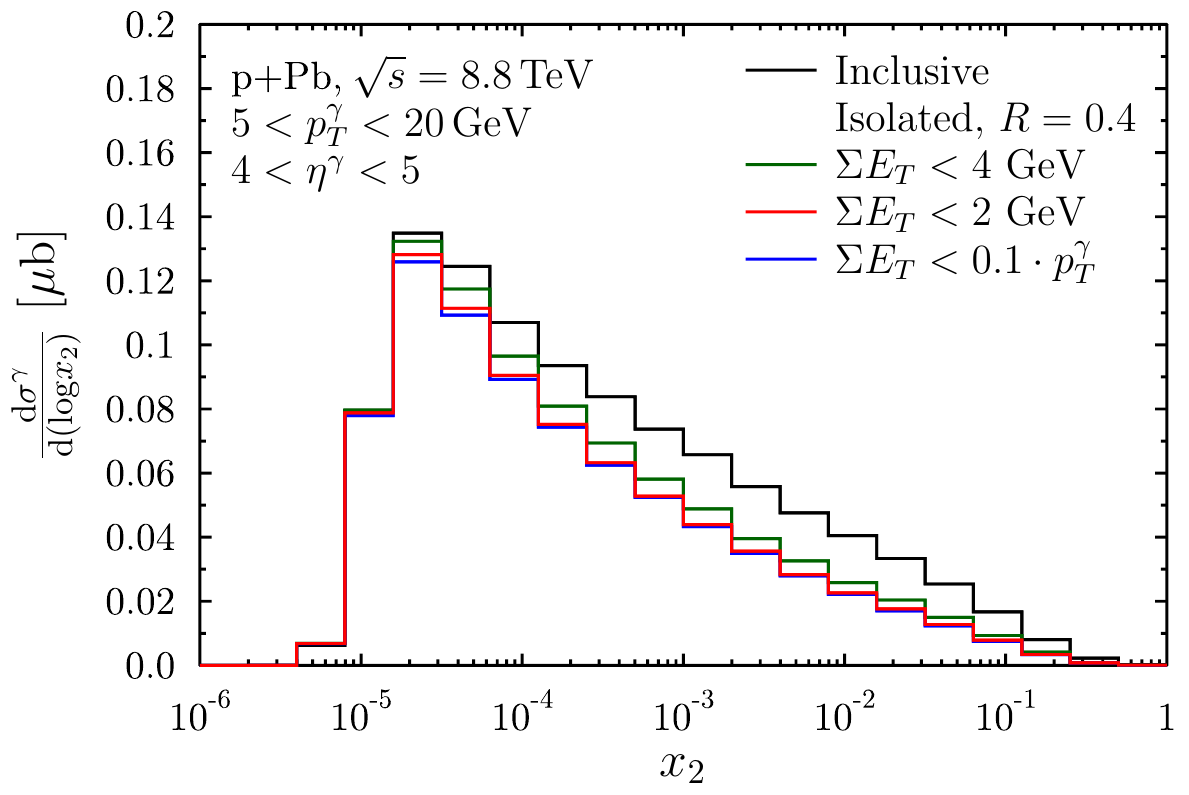}
\caption{The $x_2$ distribution for inclusive direct photons (black) and for isolated photons with $\Sigma E_T< 4\,\mathrm{GeV}$ (green), $\Sigma E_T< 2\,\mathrm{GeV}$ (red), and $\Sigma E_T< 0.1\cdot p_T$ (blue) using cone radius $R=0.4$.}
\label{fig:dsigma_x2_inc_vs_iso_gamma}
\end{minipage}
\hspace{0.02\linewidth}
\begin{minipage}[t]{0.48\linewidth}
\centering
\includegraphics[width=\textwidth]{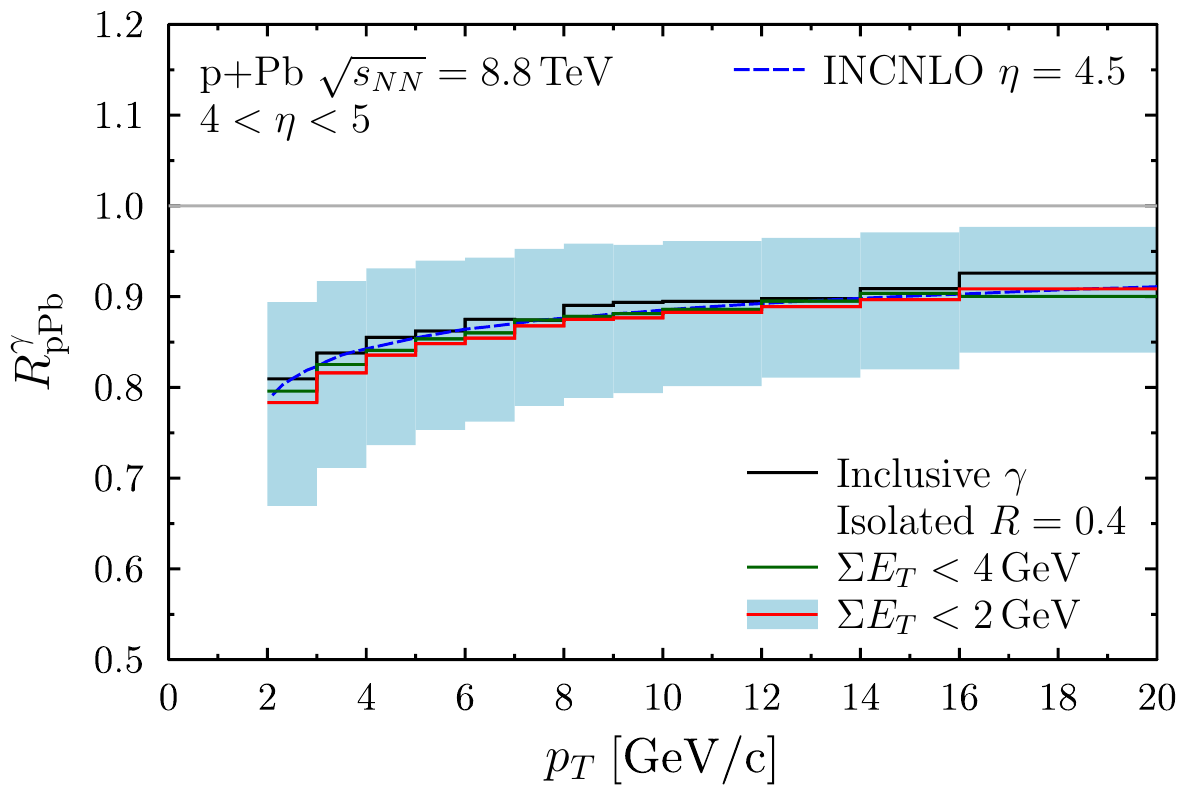}
\caption{The nuclear modification ratio $R_{\rm pPb}^{\gamma}$ for inclusive (black) and isolated photons with $\Sigma E_T< 4\,\mathrm{GeV}$ (green) and $\Sigma E_T< 2\,\mathrm{GeV}$ (red) calculated with \texttt{JETPHOX} using EPS09 nPDFs. The uncertainty band is calculated for $\Sigma E_T< 2\,\mathrm{GeV}$. Also the $R_{\rm pPb}^{\gamma}$ for the inclusive $\gamma$ from \texttt{INCNLO} is plotted (blue dashed).}
\label{fig:R_pPb_iso_gamma_y45}
\end{minipage}
\end{figure}

Figure \ref{fig:dsigma_x2_inc_vs_iso_gamma} shows the differential cross sections for inclusive photons, isolated photons with $\Sigma E_T<4\,\mathrm{GeV}$ and $\Sigma E_T<2\,\mathrm{GeV}$, and $\Sigma E_T<0.1\cdot p_T^{\gamma}$ using $R=0.4$, as a function of $x_2$. The systematics are clear: upon imposing an isolation cut $\Sigma E_T<4\,\mathrm{GeV}$ the contribution to the total cross section from larger $x_2$ values is less in comparison to the inclusive photons as the fragmentation component is suppressed. With a tighter isolation cut, $\Sigma E_T<2\,\mathrm{GeV}$, the fragmentation component is suppressed even further. Defining the upper limit of the allowed hadronic energy to be $10\,\%$ of photon $p_T$ has a very similar isolation-cut effect as the fixed limit $\Sigma E_T<2\,\mathrm{GeV}$.

Despite the increased small-$x_2$ sensitivity, the isolation cuts have only a small effect on $R^{\gamma}_{\rm pPb}$, as shown in figure \ref{fig:R_pPb_iso_gamma_y45} (which could have been anticipated already based on figures \ref{fig:dsigma_x2_inc_gamma} and \ref{fig:R_pPb_inc_gamma}). At $p_T<7\,\mathrm{GeV/c}$ only a slightly stronger suppression than in the inclusive direct photon case is observed. At larger $p_T$, the difference is easily of the same order than the numerical fluctuations arising from the limited statistics in MC sampling. To cross-check our results and the reliability of the sampling in the kinematical region studied we show, in figure \ref{fig:R_pPb_iso_gamma_y45}, also the ratio $R_{\rm pPb}^{\gamma}$ for the inclusive photons from the \texttt{INCNLO} code: The results nicely coincide with those from \texttt{JETPHOX}.
The nPDF-originating uncertainty band for the isolated photons is again computed with the error sets of EPS09 and, as expected, the error band is of the same size as for the inclusive photons in figure \ref{fig:R_pPb_inc_gamma}.

\begin{figure}[htb]
\begin{minipage}[t]{0.48\linewidth}
\centering
\includegraphics[width=\textwidth]{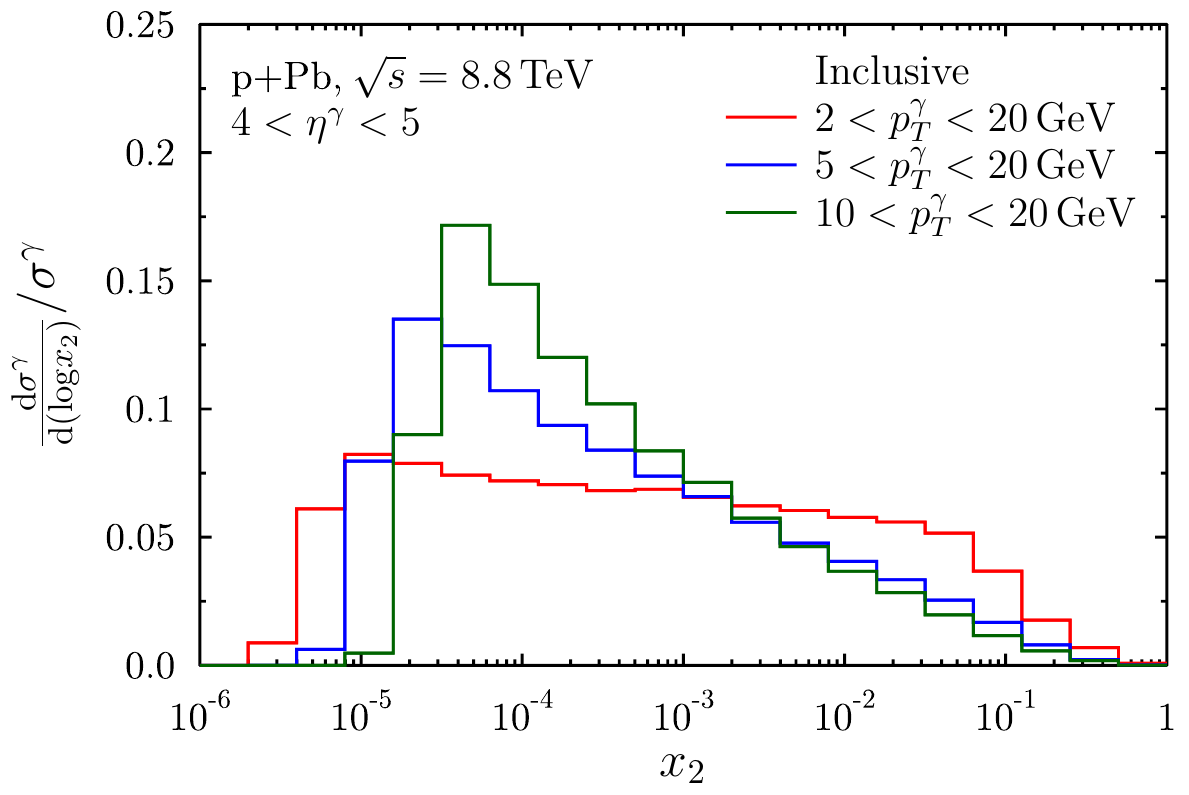}
\caption{Normalized $x_2$ distribution for inclusive $\gamma$ production at $\eta=4.5$ for $2<p_T<20\,\mathrm{GeV/c}$ (red), $5<p_T<20\,\mathrm{GeV/c}$ (blue), and $10<p_T<20\,\mathrm{GeV/c}$ (green).}
\label{fig:dsigma_x2_inc_gamma_pt}
\end{minipage}
\hspace{0.02\linewidth}
\begin{minipage}[t]{0.48\linewidth}
\centering
\includegraphics[width=\textwidth]{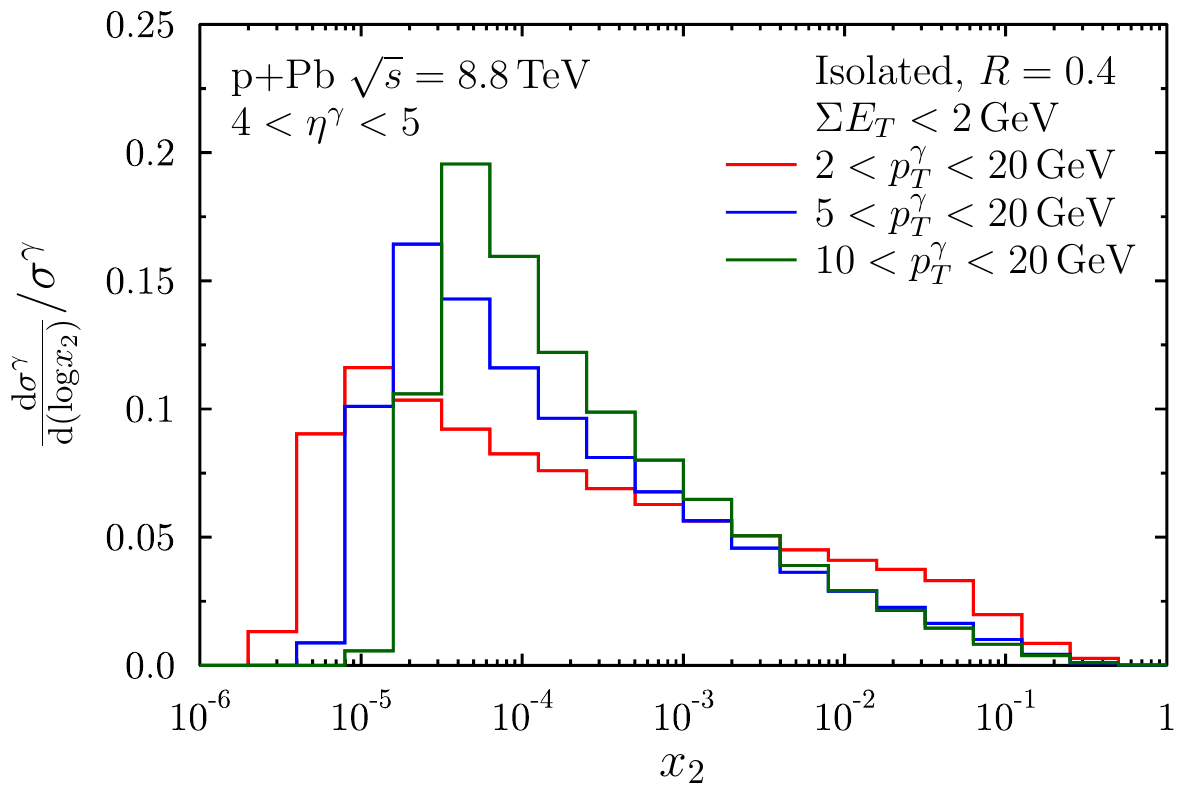}
\caption{Normalized $x_2$ distribution for isolated $\gamma$ production with $R=0.4$ and $\Sigma E_T<2\,\mathrm{GeV}$ at $4<\eta<5$ for $2<p_T<20\,\mathrm{GeV/c}$ (red), $5<p_T<20\,\mathrm{GeV/c}$ (blue), and $10<p_T<20\,\mathrm{GeV/c}$ (green).}
\label{fig:dsigma_x2_iso2_gamma_pt}
\end{minipage}
\end{figure}

To study the effect of an isolation cut in different $p_T$ regions, the normalized $x_2$ distribution of the inclusive photon cross section is plotted in figure \ref{fig:dsigma_x2_inc_gamma_pt} with three different lower limits of $p_T^{\gamma}$, $2, 5, 10\,\mathrm{GeV/c}$, and in figure \ref{fig:dsigma_x2_iso2_gamma_pt} for isolated photons with $\Sigma E_T<2\,\mathrm{GeV}$. Similarly as for $\pi^0$'s above, pushing the calculation down to $p_T\sim 2\,\mathrm{GeV/c}$ actually increases the contribution from the $x_2>0.01$ region which corresponds to the antishadowing region in the EPS09 nPDFs. The isolation cut suppresses the tail at large $x_2$ which is  not a dramatic effect but explains the slightly stronger suppression of $R_{\rm pPb}^{\gamma}$ at low $p_T$.

\begin{figure}[htb]
\begin{minipage}[t]{0.48\linewidth}
\centering
\includegraphics[width=\textwidth]{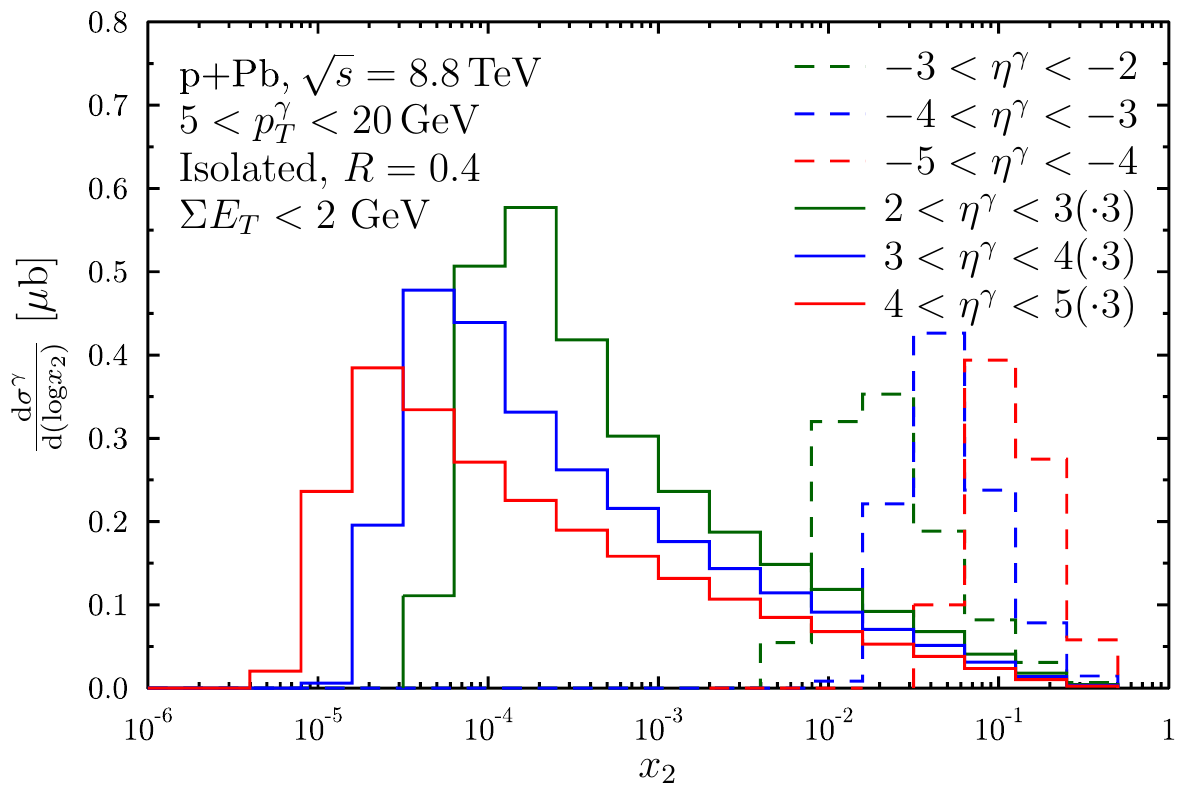}
\caption{The absolute $x_2$ distribution for isolated $\gamma$ production with $R=0.4$ and $\Sigma E_T<2\,\mathrm{GeV}$ at $2<p_T<20\,\mathrm{GeV/c}$ for $2<\eta<3$ (green), $3<\eta<4$ (blue), and $4<\eta<5$ (red) at forward (solid) and corresponding backward (dashed) rapidities. Note that the forward-$\eta$ distributions have been multiplied by a factor 3 for a better readability.}
\label{fig:dsigma_x2_iso_gamma}
\end{minipage}
\hspace{0.02\linewidth}
\begin{minipage}[t]{0.48\linewidth}
\centering
\includegraphics[width=\textwidth]{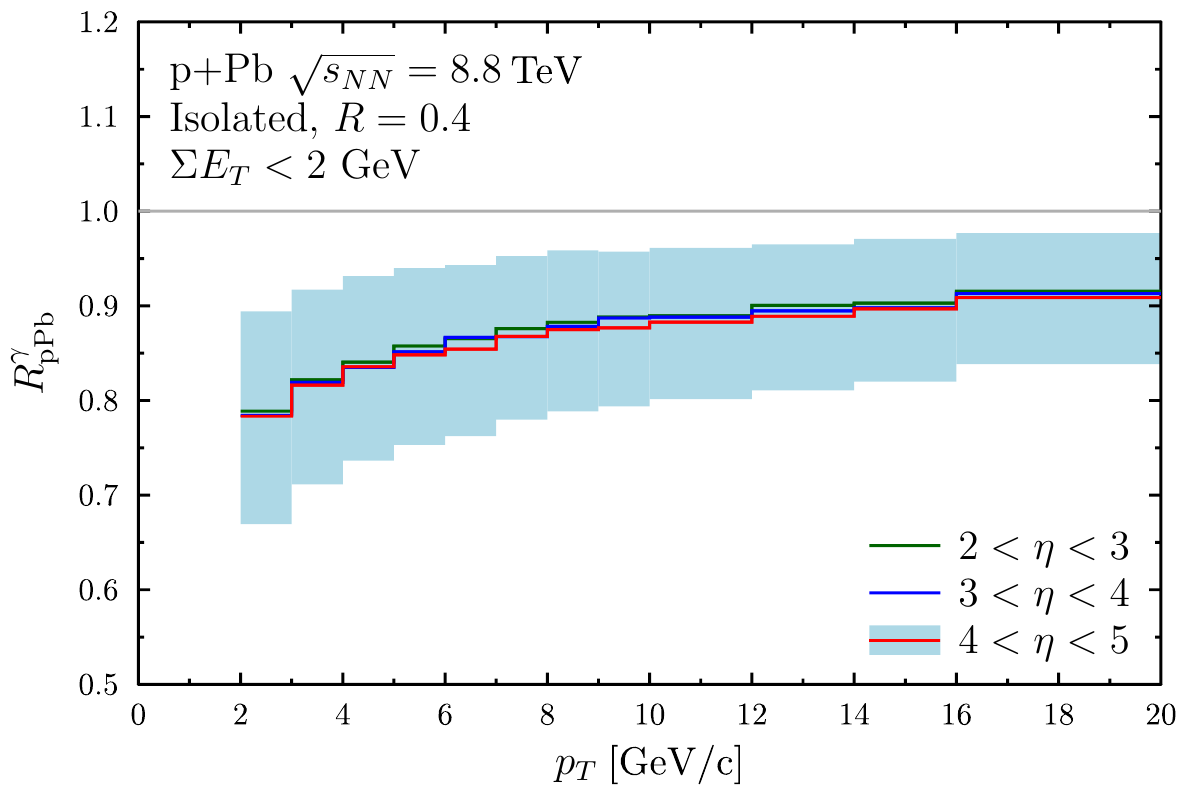}
\caption{The nuclear modification ratio $R_{\rm pPb}^{\gamma}$ for isolated $\gamma$ production with $R=0.4$ and $\Sigma E_T<2\,\mathrm{GeV}$ for rapidities  $2<\eta<3$ (green), $3<\eta<4$ (blue), and $4<\eta<5$ (red). The blue EPS09 uncertainty band is calculated for $4<\eta<5$.}
\label{fig:R_pPb_gamma_y4}
\end{minipage}
\end{figure}

To check the expected rapidity systematics of the nucelar effects in direct photon production with isolation cuts we plot, in figure \ref{fig:dsigma_x2_iso_gamma}, the $x_2$ distribution of the cross section $\mathrm{d} \sigma^{\gamma+X}_{\rm pPb}/dp_Td\eta$ at different forward-rapidity bins integrated over $5<p_T<20$~GeV/c. For the discussion presented in the next subsection, also the $x_2$ distributions at backward rapidities are shown. The isolation cuts have reduced the fragmentation tails at larger $x_2$, and made the cross sections somewhat more sensitive to the small-$x_2$ region. Towards more forward rapidities the probed values of $x_2$ decrease but as the DGLAP evolution quickly washes out all strong effects from small-$x$ gluons the ratio $R_{\rm pPb}^{\gamma}$, presented in figure \ref{fig:R_pPb_gamma_y4}, shows practically no rapidity dependence at forward direction. This suggests that --- as far as nuclear effects in PDFs are concerned --- there is no practical advantage for considering observables at very forward rapidity ($4<\eta<5$) but moderate values ($2<\eta<3$) for which the cross sections are larger should be sufficient. On the other hand, an observation of a clearly stronger $R_{\rm pPb}^{\gamma}$ rapidity dependence could be a signature of physics beyond the DGLAP framework.

\subsection{Forward-to-backward ratio}

\begin{figure}[htb]
\begin{minipage}[t]{0.48\linewidth}
\centering
\includegraphics[width=\textwidth]{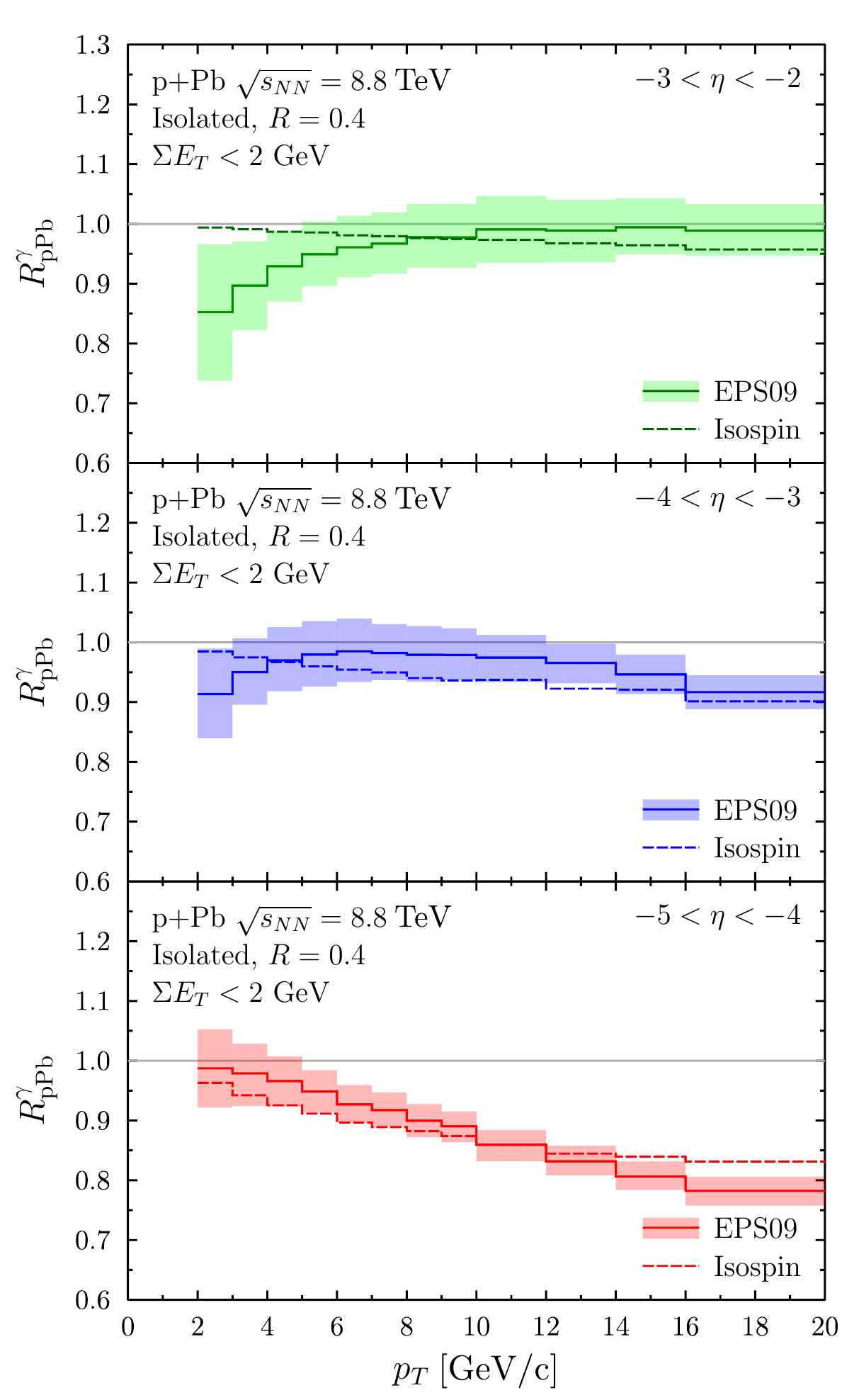}
\caption{The nuclear modification $R_{\rm pPb}^{\gamma}$ for isolated $\gamma$ production with $R=0.4$ and $\Sigma E_T<2\,\mathrm{GeV}$ for rapidities  $-3<\eta<-2$ (top), $-4<\eta<-3$ (middle), and $-5<\eta<-4$ (bottom). The error bands are from EPS09. The isospin effect (dashed) is also shown.}
\label{fig:R_pPb_back}
\end{minipage}
\hspace{0.02\linewidth}
\begin{minipage}[t]{0.48\linewidth}
\centering
\includegraphics[width=\textwidth]{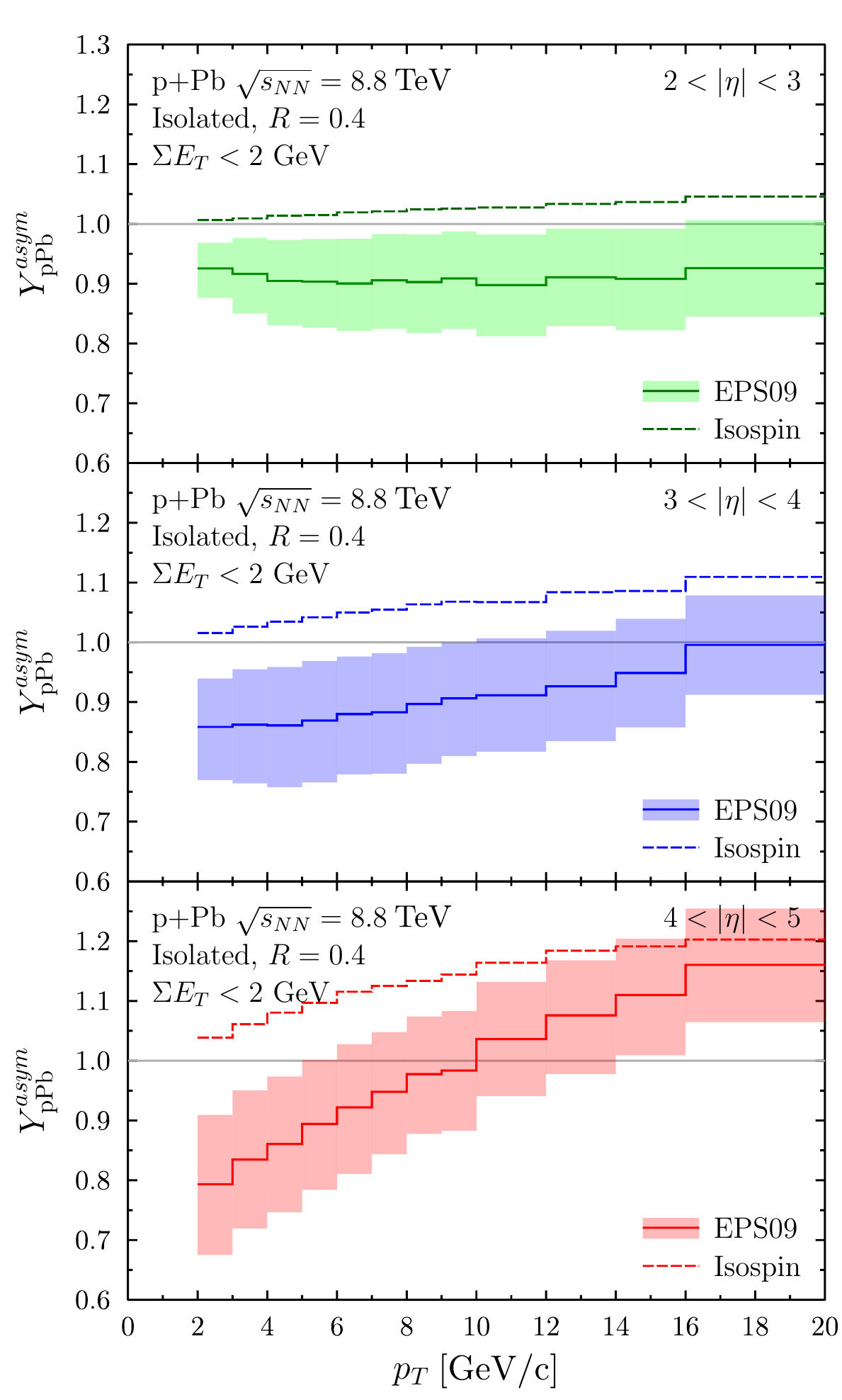}
\caption{The yield asymmetry $Y_{asym}^{\gamma}$ for isolated $\gamma$ production with $R=0.4$ and $\Sigma E_T<2\,\mathrm{GeV}$ for rapidities  $2<|\eta|<3$ (top), $3<|\eta|<4$ (middle), and $4<|\eta|<5$ (bottom). The error bands are from EPS09. The isospin effect (dashed) is also shown.}
\label{fig:R_fb}
\end{minipage}
\end{figure}

Until now, we have used solely the nuclear modification ratios $R_{\rm pPb}^{\gamma}$ to quantify the nuclear effects and the calculated EPS09 error bands suggest that to obtain significant further constraints, one should be able to measure $R_{\rm pPb}^{\gamma}$ with better than a $10$\% precision.  If there is no p+p baseline measurement with the same $\sqrt{s_{NN}}$ available, this may be very challenging. Also, if the luminosity for the collected data sample is not measured, the conversion from the measured yields to cross sections may involve Glauber modeling \cite{Miller:2007ri} causing some overall normalization uncertainty whose implementation in a $\chi^2$ analysis is somewhat ambiguous. It would obviously be preferable to consider observables that are free from such uncertainties. An option that has already been recognized useful in p+Pb collisions (see e.g. \cite{CMS:2013cka,Abelev:2013yxa}) is the to form the yield asymmetry between the forward and backward rapidities,
\begin{equation}
Y_{\rm pPb}^{\rm asym} \equiv Y_{\rm pPb}^{\rm asym}(p_T,\eta) \equiv \left.\frac{\mathrm{d}^2\sigma_{\rm pPb}}{\mathrm{d}p_T \mathrm{d}\eta}\right|_{\eta\in[\eta_1,\eta_2]}\bigg/ \left.\frac{\mathrm{d}^2\sigma_{\rm pPb}}{\mathrm{d}p_T \mathrm{d}\eta}\right|_{\eta\in[-\eta_2,-\eta_1]}.
\label{eq:Yasym}
\end{equation}
In addition to being free from the absolute normalization uncertainty some correlated systematic uncertainties can be expected to cancel as well (in a similar fashion as jet energy-scale uncertainties largely cancel in ratios of inclusive jet cross sections between different $\sqrt{s}$ but fixed rapidity and $p_T$ \cite{Aad:2013lpa}).
As indicated by the dashed lines in figure \ref{fig:dsigma_x2_iso_gamma}, the isolated photon production at backward rapidities will be sensitive to the region $0.01<x_2<0.2$, which corresponds to antishadowing and EMC effect in EPS09. This kinematic region starts to be sensitive also to the nuclear valence quarks giving rise to an ''isospin effect``, which follows from the lower charge density of neutrons in comparison to the protons. As the photon cross sections are proportional to the electromagnetic charge, some suppression due to the presence of neutrons in the nucleus is expected. These effects can be easily quantified by calculating the $R_{\rm pPb}^{\gamma}$ without the nuclear modifications in the PDFs and are shown by the dashed lines in figure \ref{fig:R_pPb_back} at rapidities $\eta \in [-5,-4]$, $[-4,-3]$, and $[-3,-2]$. Indeed, the isospin effect becomes prominent for $\eta < -3$. The expected total nuclear modifications are shown with the EPS09 error bands. As there are already other data constraints at large $x_2$, the nPDF errors are clearly smaller than the corresponding bands in the forward direction. With different rapidity bins we observe different effects: at $-3 < \eta < -2$ and $-4 < \eta < -3$ we have first some suppression in comparison to the isospin baseline which eventually turns to an enhancement caused by the antishadowing in EPS09. At $-5 < \eta < -4$ the isolated photons are already sensitive to the EMC region at $p_T>12\,\mathrm{GeV/c}$. In general the nPDF-originating uncertainties appear smaller than $5\,\%$ except at $p_T<5\,\mathrm{GeV/c}$ in $-3 < \eta < -2$ bin, which indicates that the isolated photon production at backward rapidities could be a better baseline option to resolve the small-$x$ effects than the p+p. However, it is not clear to what extent the unknown flavor dependence of the nuclear effects in PDFs affects this situation.

The forward-to-backward yield asymmetries $Y_{\rm pPb}^{\rm asym}$ for the isolated photon production are plotted in figure \ref{fig:R_fb}, again for the three rapidity bins, $|\eta| \in [2,3]$, $[3,4]$, and $[4,5]$ with the PDF nuclear modifications and their uncertainties. The results including only the isospin effect are shown for comparison. The uncertainty bands have been computed by forming the observable $Y_{\rm pPb}^{\rm asym}$ with each of the EPS09 error sets first, and computing the error band as instructed in \cite{Eskola:2009uj}. Then, if the forward and backward regions are sensitive to the same nuclear effect, such as shadowing at small $p_T$ in the bin $2 < |\eta| < 3$, there is a partial cancellation of the uncertainties in $Y_{\rm pPb}^{\rm asym}$ -- see the small-$p_T$ region of the first panel. Mostly, however, the yield asymmetries are sensitive to two very different $x_2$ regions and the uncertainties add up. As the isolated photon ratios $R_{\rm pPb}^{\gamma}$ at different forward rapidities (figure \ref{fig:R_pPb_gamma_y4}) are very similar, the significant rapidity dependence of $Y_{\rm pPb}^{\rm asym}$ (for fixed $p_T$) follows mostly from the nuclear effects at backward rapidities (figure \ref{fig:R_pPb_back}). The larger nPDF uncertainties for $R_{\rm pPb}^{\gamma}$ at forward rapidities than at backward direction suggest that the theoretical uncertainties in $Y_{\rm pPb}^{\rm asym}$ are mostly due to the lack of nPDF constraints at small $x_2$, and measurements of $Y_{\rm pPb}^{\rm asym}$ with sufficient accuracy would improve this situation. At least, the predicted total effect is large and thus the yield asymmetry $Y_{\rm pPb}^{\rm asym}$ would in any case serve as a further test of the collinear factorization and e.g. the treatment of isospin effects in nuclear collisions.

\section{Conclusions}

We have studied inclusive direct photon production at forward rapidities in p+Pb collisions at the LHC, trying to sort out the $x_2$ regions that could be probed by measuring them at different kinematic corners. We have shown that, for fixed kinematics in the forward direction, the direct photons are sensitive to smaller values of $x_2$ than the inclusive hadrons and by imposing an isolation cut for the direct photons one can increase such sensitivity even more. It turns out that the naive LO-based kinematics are a rather poor estimate when it comes to finding the predominantly important $x_2$ regions and that full NLO calculations are needed in order to understand the true widths and shapes of these distributions. In particular, at forward rapidities the cross sections are affected by a wide range of $x_2$ values. This is true especially at low $p_T$ and --- a little bit counterintuitively --- we find the observables around $p_T \sim 5\,\mathrm{GeV/c}$ to be actually more sensitive to small-$x_2$ partons than the same observables at even lower $p_T$. The expected nuclear modifications $R_{\rm pPb}^{\gamma}$ are, however, found to be almost completely insensitive to whether an isolation cut is applied or not and practically independent of the rapidity beyond $\eta > 2$. The main reason for such behaviour is the DGLAP evolution of the gluon PDFs which rapidly smooths out any strong nuclear effects in gluon PDFs. As an alternative to the canonical $R_{\rm pPb}^{\gamma}$, we have considered also the yield asymmetry between the forward and backward rapidities which does not require a measurement of the p+p baseline and could presumably be measured with a better accuracy than $R_{\rm pPb}^{\gamma}$.

Even though our focus here has been on the gluon nPDFs and all our calculations rely on the collinear factorization and linear DGLAP dynamics, the importance of direct photons as a probe of possible deviations from this standard theoretical framework should not be forgotten. As the non-linearities are foreseen to play a role at sufficiently small $x_2$, a systematic search at the LHC forward rapidities would lead to a better understanding concerning the onset of such phenomena.

\acknowledgments
We acknowledge the financial support from the Magnus Ehrnrooth Foundation (I.H.) and from the Academy of Finland, Project No. 133005. We thank Thomas Peitzmann and Marco van Leeuwen for helpful discussions. Related to this work, I.H. thanks the ALICE group at the Utrecht University for hospitality and discussions during his visit there.

\bibliographystyle{JHEP}
\bibliography{iso_phot}

\providecommand{\href}[2]{#2}\begingroup\raggedright\begin{thebibliography}{10}

\bibitem{Collins:1989gx}
J.~C. Collins, D.~E. Soper, and G.~F. Sterman, {\it {Factorization of Hard
  Processes in QCD}},  {\em Adv.~Ser.~Direct.~High Energy Phys.} {\bf 5} (1988)
  1--91, [\href{http://xxx.lanl.gov/abs/hep-ph/0409313}{{\tt hep-ph/0409313}}].

\bibitem{Brock:1993sz}
{\bf CTEQ} Collaboration, R.~Brock et~al., {\it {Handbook of perturbative QCD:
  Version 1.0}},  {\em Rev.~Mod.~Phys.} {\bf 67} (1995) 157--248.

\bibitem{Forte:2013wc}
S.~Forte and G.~Watt, {\it {Progress in the Determination of the Partonic
  Structure of the Proton}},  {\em Ann.~Rev.~Nucl.~Part.~Sci.} {\bf 63} (2013)
  291--328, [\href{http://xxx.lanl.gov/abs/1301.6754}{{\tt arXiv:1301.6754}}].

\bibitem{Eskola:2009uj}
K.~J. Eskola, H.~Paukkunen, and C.~A. Salgado, {\it {EPS09 - a New Generation
  of NLO and LO Nuclear Parton Distribution Functions}},  {\em JHEP} {\bf 04}
  (2009) 065, [\href{http://xxx.lanl.gov/abs/0902.4154}{{\tt
  arXiv:0902.4154}}].

\bibitem{Hirai:2007sx}
M.~Hirai, S.~Kumano, and T.-H. Nagai, {\it {Determination of nuclear parton
  distribution functions and their uncertainties in next-to-leading order}},
  {\em Phys.~Rev.} {\bf C76} (2007) 065207,
  [\href{http://xxx.lanl.gov/abs/0709.3038}{{\tt arXiv:0709.3038}}].

\bibitem{deFlorian:2011fp}
D.~de~Florian, R.~Sassot, P.~Zurita, and M.~Stratmann, {\it {Global Analysis of
  Nuclear Parton Distributions}},  {\em Phys.~Rev.} {\bf D85} (2012) 074028,
  [\href{http://xxx.lanl.gov/abs/1112.6324}{{\tt arXiv:1112.6324}}].

\bibitem{Schienbein:2009kk}
I.~Schienbein, J.~Y. Yu, K.~Kovarik, C.~Keppel, J.~Morfin, et~al., {\it {PDF
  Nuclear Corrections for Charged and Neutral Current Processes}},  {\em
  Phys.~Rev.} {\bf D80} (2009) 094004,
  [\href{http://xxx.lanl.gov/abs/0907.2357}{{\tt arXiv:0907.2357}}].

\bibitem{Kovarik:2013sya}
K.~Kovarik, T.~Jezo, A.~Kusina, F.~I. Olness, I.~Schienbein, et~al., {\it {CTEQ
  nuclear parton distribution functions}},  {\em PoS} {\bf DIS2013} (2013) 274,
  [\href{http://xxx.lanl.gov/abs/1307.3454}{{\tt arXiv:1307.3454}}].

\bibitem{Eskola:2012rg}
K.~J. Eskola, {\it {Global analysis of nuclear PDFs - latest developments}},
  {\em Nucl.~Phys.} {\bf A 910-911} (2013) 163,
  [\href{http://xxx.lanl.gov/abs/1209.1546}{{\tt arXiv:1209.1546}}].

\bibitem{Paukkunen:2014nqa}
H.~Paukkunen, {\it {Nuclear PDFs in the beginning of the LHC era}},
  \href{http://xxx.lanl.gov/abs/1401.2345}{{\tt arXiv:1401.2345}}.

\bibitem{Eskola:1998df}
K.~J. Eskola, V.~J. Kolhinen, and C.~A. Salgado, {\it {The Scale dependent
  nuclear effects in parton distributions for practical applications}},  {\em
  Eur.~Phys.~J.} {\bf C9} (1999) 61--68,
  [\href{http://xxx.lanl.gov/abs/hep-ph/9807297}{{\tt hep-ph/9807297}}].

\bibitem{Lipatov:1974qm}
L.~N. Lipatov, {\it {The parton model and perturbation theory}},  {\em
  Sov.~J.~Nucl.~Phys.} {\bf 20} (1975) 94--102.

\bibitem{Gribov:1972ri}
V.~N. Gribov and L.~N. Lipatov, {\it {Deep inelastic e p scattering in
  perturbation theory}},  {\em Sov.~J.~Nucl.~Phys.} {\bf 15} (1972) 438--450.

\bibitem{Altarelli:1977zs}
G.~Altarelli and G.~Parisi, {\it {Asymptotic Freedom in Parton Language}},
  {\em Nucl.~Phys.} {\bf B126} (1977) 298.

\bibitem{Dokshitzer:1977sg}
Y.~L. Dokshitzer, {\it {Calculation of the Structure Functions for Deep
  Inelastic Scattering and e+ e- Annihilation by Perturbation Theory in Quantum
  Chromodynamics.}},  {\em Sov.~Phys.~JETP} {\bf 46} (1977) 641--653.

\bibitem{Pumplin:2001ct}
J.~Pumplin, D.~Stump, R.~Brock, D.~Casey, J.~Huston, et~al., {\it
  {Uncertainties of predictions from parton distribution functions. 2. The
  Hessian method}},  {\em Phys.~Rev.} {\bf D65} (2001) 014013,
  [\href{http://xxx.lanl.gov/abs/hep-ph/0101032}{{\tt hep-ph/0101032}}].

\bibitem{Paukkunen:2014zia}
H.~Paukkunen and P.~Zurita, {\it {PDF reweighting in the Hessian matrix
  approach}},  \href{http://xxx.lanl.gov/abs/1402.6623}{{\tt arXiv:1402.6623}}.

\bibitem{Chatrchyan:2014hqa}
{\bf CMS} Collaboration, S.~Chatrchyan et~al., {\it {Studies of dijet
  pseudorapidity distributions and transverse momentum balance in pPb
  collisions at $\sqrt{s_{NN}}$=5.02 TeV}},
  \href{http://xxx.lanl.gov/abs/1401.4433}{{\tt arXiv:1401.4433}}.

\bibitem{Paukkunen:2013hbm}
{\bf LHeC Study Group} Collaboration, H.~Paukkunen, K.~J. Eskola, and
  N.~Armesto, {\it {Nuclear PDFs from the LHeC perspective}},  {\em PoS} {\bf
  DIS2013} (2013) 276, [\href{http://xxx.lanl.gov/abs/1306.2486}{{\tt
  arXiv:1306.2486}}].

\bibitem{Kang:2012am}
Z.-B. Kang and J.-W. Qiu, {\it {Nuclear modification of vector boson production
  in proton-lead collisions at the LHC}},  {\em Phys.~Lett.} {\bf B721} (2013)
  277--283, [\href{http://xxx.lanl.gov/abs/1212.6541}{{\tt arXiv:1212.6541}}].

\bibitem{Brandt:2014vva}
M.~Brandt, M.~Klasen, and F.~König, {\it {Nuclear parton density modifications
  from low-mass lepton pair production at the LHC}},
  \href{http://xxx.lanl.gov/abs/1401.6817}{{\tt arXiv:1401.6817}}.

\bibitem{Guzey:2012jp}
V.~Guzey, M.~Guzzi, P.~M. Nadolsky, M.~Strikman, and B.~Wang, {\it {Massive
  neutral gauge boson production as a probe of nuclear modifications of parton
  distributions at the LHC}},  {\em Eur.~Phys.~J.} {\bf A49} (2013) 35,
  [\href{http://xxx.lanl.gov/abs/1212.5344}{{\tt arXiv:1212.5344}}].

\bibitem{Paukkunen:2010qg}
H.~Paukkunen and C.~A. Salgado, {\it {Constraints for the nuclear parton
  distributions from Z and W production at the LHC}},  {\em JHEP} {\bf 1103}
  (2011) 071, [\href{http://xxx.lanl.gov/abs/1010.5392}{{\tt
  arXiv:1010.5392}}].

\bibitem{Eskola:2013aya}
K.~J. Eskola, H.~Paukkunen, and C.~A. Salgado, {\it {A perturbative QCD study
  of dijets in p+Pb collisions at the LHC}},  {\em JHEP} {\bf 1310} (2013) 213,
  [\href{http://xxx.lanl.gov/abs/1308.6733}{{\tt arXiv:1308.6733}}].

\bibitem{Arleo:2004gn}
F.~Arleo, P.~Aurenche, F.~W. Bopp, I.~Dadic, G.~David, et~al., {\it {Hard
  probes in heavy-ion collisions at the LHC: Photon physics in heavy ion
  collisions at the LHC}},  \href{http://xxx.lanl.gov/abs/hep-ph/0311131}{{\tt
  hep-ph/0311131}}.

\bibitem{Arleo:2011gc}
F.~Arleo, K.~J. Eskola, H.~Paukkunen, and C.~A. Salgado, {\it {Inclusive prompt
  photon production in nuclear collisions at RHIC and LHC}},  {\em JHEP} {\bf
  1104} (2011) 055, [\href{http://xxx.lanl.gov/abs/1103.1471}{{\tt
  arXiv:1103.1471}}].

\bibitem{Stavreva:2010mw}
T.~Stavreva, I.~Schienbein, F.~Arleo, K.~Kovarik, F.~Olness, et~al., {\it
  {Probing gluon and heavy-quark nuclear PDFs with gamma + Q production in pA
  collisions}},  {\em JHEP} {\bf 1101} (2011) 152,
  [\href{http://xxx.lanl.gov/abs/1012.1178}{{\tt arXiv:1012.1178}}].

\bibitem{Dai:2013xca}
W.~Dai, S.-Y. Chen, B.-W. Zhang, and E.-K. Wang, {\it {Cold nuclear matter
  effects on isolated prompt photon and isolated prompt photon+jet productions
  in relativistic heavy-ion collisions}},  {\em Commun.~Theor.~Phys.} {\bf 59}
  (2013) 349--355.

\bibitem{Arleo:2007js}
F.~Arleo and T.~Gousset, {\it {Measuring gluon shadowing with prompt photons at
  RHIC and LHC}},  {\em Phys.~Lett.} {\bf B660} (2008) 181--187,
  [\href{http://xxx.lanl.gov/abs/0707.2944}{{\tt arXiv:0707.2944}}].

\bibitem{Albacete:2013ei}
J.~L. Albacete, N.~Armesto, R.~Baier, G.~G. Barnafoldi, J.~Barrette, et~al.,
  {\it {Predictions for $p+$Pb Collisions at sqrt s\_NN = 5 TeV}},  {\em
  Int.~J.~Mod.~Phys.} {\bf E22} (2013) 1330007,
  [\href{http://xxx.lanl.gov/abs/1301.3395}{{\tt arXiv:1301.3395}}].

\bibitem{Helenius:2013bya}
I.~Helenius, K.~J. Eskola, and H.~Paukkunen, {\it {Centrality dependence of
  inclusive prompt photon production in d+Au, Au+Au, p+Pb, and Pb+Pb
  collisions}},  {\em JHEP} {\bf 1305} (2013) 030,
  [\href{http://xxx.lanl.gov/abs/1302.5580}{{\tt arXiv:1302.5580}}].

\bibitem{d'Enterria:2012yj}
D.~d'Enterria and J.~Rojo, {\it {Quantitative constraints on the gluon
  distribution function in the proton from collider isolated-photon data}},
  {\em Nucl.~Phys.} {\bf B860} (2012) 311--338,
  [\href{http://xxx.lanl.gov/abs/1202.1762}{{\tt arXiv:1202.1762}}].

\bibitem{Mueller:1985wy}
A.~H. Mueller and J.-w. Qiu, {\it {Gluon Recombination and Shadowing at Small
  Values of x}},  {\em Nucl.~Phys.} {\bf B268} (1986) 427.

\bibitem{Eskola:2002yc}
K.~J. Eskola, H.~Honkanen, V.~J. Kolhinen, J.-w. Qiu, and C.~A. Salgado, {\it
  {Nonlinear corrections to the DGLAP equations in view of the hera data}},
  {\em Nucl.~Phys.} {\bf B660} (2003) 211--224,
  [\href{http://xxx.lanl.gov/abs/hep-ph/0211239}{{\tt hep-ph/0211239}}].

\bibitem{Gribov:1984tu}
L.~V. Gribov, E.~M. Levin, and M.~G. Ryskin, {\it {Semihard Processes in QCD}},
   {\em Phys.~Rept.} {\bf 100} (1983) 1--150.

\bibitem{McLerran:1993ni}
L.~D. McLerran and R.~Venugopalan, {\it {Computing quark and gluon distribution
  functions for very large nuclei}},  {\em Phys.~Rev.} {\bf D49} (1994)
  2233--2241, [\href{http://xxx.lanl.gov/abs/hep-ph/9309289}{{\tt
  hep-ph/9309289}}].

\bibitem{Albacete:2014fwa}
J.~L. Albacete and C.~Marquet, {\it {Gluon saturation and initial conditions
  for relativistic heavy ion collisions}},  {\em Prog.~Part.~Nucl.~Phys.} {\bf
  76} (2014) 1--42, [\href{http://xxx.lanl.gov/abs/1401.4866}{{\tt
  arXiv:1401.4866}}].

\bibitem{JalilianMarian:2012bd}
J.~Jalilian-Marian and A.~H. Rezaeian, {\it {Prompt photon production and
  photon-hadron correlations at RHIC and the LHC from the Color Glass
  Condensate}},  {\em Phys.~Rev.} {\bf D86} (2012) 034016,
  [\href{http://xxx.lanl.gov/abs/1204.1319}{{\tt arXiv:1204.1319}}].

\bibitem{Rezaeian:2012ye}
A.~H. Rezaeian, {\it {CGC predictions for p+A collisions at the LHC and
  signature of QCD saturation}},  {\em Phys.~Lett.} {\bf B718} (2013)
  1058--1069, [\href{http://xxx.lanl.gov/abs/1210.2385}{{\tt
  arXiv:1210.2385}}].

\bibitem{T.PeitzmannfortheALICEFoCal:2013fja}
{\bf ALICE} Collaboration, T.~Peitzmann, {\it {Prototype studies for a forward
  EM calorimeter in ALICE}},  \href{http://xxx.lanl.gov/abs/1308.2585}{{\tt
  arXiv:1308.2585}}.

\bibitem{incnlopage}
\url{http://lapth.in2p3.fr/PHOX_FAMILY/readme_inc.html}.

\bibitem{Aversa:1988vb}
F.~Aversa, P.~Chiappetta, M.~Greco, and J.~P. Guillet, {\it {QCD Corrections to
  Parton-Parton Scattering Processes}},  {\em Nucl.~Phys.} {\bf B327} (1989)
  105.

\bibitem{Aurenche:1987fs}
P.~Aurenche, R.~Baier, M.~Fontannaz, and D.~Schiff, {\it {Prompt Photon
  Production at Large p(T) Scheme Invariant QCD Predictions and Comparison with
  Experiment}},  {\em Nucl.~Phys.} {\bf B297} (1988) 661.

\bibitem{Aurenche:1998gv}
P.~Aurenche, M.~Fontannaz, J.~P. Guillet, B.~A. Kniehl, E.~Pilon, et~al., {\it
  {A Critical phenomenological study of inclusive photon production in hadronic
  collisions}},  {\em Eur.~Phys.~J.} {\bf C9} (1999) 107--119,
  [\href{http://xxx.lanl.gov/abs/hep-ph/9811382}{{\tt hep-ph/9811382}}].

\bibitem{Aurenche:1999nz}
P.~Aurenche, M.~Fontannaz, J.~P. Guillet, B.~A. Kniehl, and M.~Werlen, {\it
  {Large p(T) inclusive pi0 cross-sections and next-to-leading-order QCD
  predictions}},  {\em Eur.~Phys.~J.} {\bf C13} (2000) 347--355,
  [\href{http://xxx.lanl.gov/abs/hep-ph/9910252}{{\tt hep-ph/9910252}}].

\bibitem{deFlorian:2007aj}
D.~de~Florian, R.~Sassot, and M.~Stratmann, {\it {Global analysis of
  fragmentation functions for pions and kaons and their uncertainties}},  {\em
  Phys.~Rev.} {\bf D75} (2007) 114010,
  [\href{http://xxx.lanl.gov/abs/hep-ph/0703242}{{\tt hep-ph/0703242}}].

\bibitem{Nadolsky:2008zw}
P.~M. Nadolsky, H.-L. Lai, Q.-H. Cao, J.~Huston, J.~Pumplin, et~al., {\it
  {Implications of CTEQ global analysis for collider observables}},  {\em
  Phys.~Rev.} {\bf D78} (2008) 013004,
  [\href{http://xxx.lanl.gov/abs/0802.0007}{{\tt arXiv:0802.0007}}].

\bibitem{Eskola:2002kv}
K.~J. Eskola and H.~Honkanen, {\it {A Perturbative QCD analysis of charged
  particle distributions in hadronic and nuclear collisions}},  {\em
  Nucl.~Phys.} {\bf A713} (2003) 167--187,
  [\href{http://xxx.lanl.gov/abs/hep-ph/0205048}{{\tt hep-ph/0205048}}].

\bibitem{Sassot:2010bh}
R.~Sassot, P.~Zurita, and M.~Stratmann, {\it {Inclusive Hadron Production in
  the CERN-LHC Era}},  {\em Phys.~Rev.} {\bf D82} (2010) 074011,
  [\href{http://xxx.lanl.gov/abs/1008.0540}{{\tt arXiv:1008.0540}}].

\bibitem{d'Enterria:2013vba}
D.~d'Enterria, K.~J. Eskola, I.~Helenius, and H.~Paukkunen, {\it {Confronting
  current NLO parton fragmentation functions with inclusive charged-particle
  spectra at hadron colliders}},  {\em Nucl.~Phys.} {\bf B883} (2014) 615--628,
  [\href{http://xxx.lanl.gov/abs/1311.1415}{{\tt arXiv:1311.1415}}].

\bibitem{Guzey:2004zp}
V.~Guzey, M.~Strikman, and W.~Vogelsang, {\it {Observations on dA scattering at
  forward rapidities}},  {\em Phys.~Lett.} {\bf B603} (2004) 173--183,
  [\href{http://xxx.lanl.gov/abs/hep-ph/0407201}{{\tt hep-ph/0407201}}].

\bibitem{Chatterjee:2013naa}
R.~Chatterjee, H.~Holopainen, I.~Helenius, T.~Renk, and K.~J. Eskola, {\it
  {Elliptic flow of thermal photons from event-by-event hydrodynamic model}},
  {\em Phys.~Rev.} {\bf C88} (2013) 034901,
  [\href{http://xxx.lanl.gov/abs/1305.6443}{{\tt arXiv:1305.6443}}].

\bibitem{Klasen:2013mga}
M.~Klasen, C.~Klein-Bösing, F.~König, and J.~P. Wessels, {\it {How robust is
  a thermal photon interpretation of the ALICE low-$p_{T}$ data?}},  {\em JHEP}
  {\bf 1310} (2013) 119, [\href{http://xxx.lanl.gov/abs/1307.7034}{{\tt
  arXiv:1307.7034}}].

\bibitem{jetphoxpage}
\url{http://lapth.cnrs.fr/PHOX_FAMILY/jetphox.html}.

\bibitem{Catani:2002ny}
S.~Catani, M.~Fontannaz, J.-P. Guillet, and E.~Pilon, {\it {Cross-section of
  isolated prompt photons in hadron hadron collisions}},  {\em JHEP} {\bf 0205}
  (2002) 028, [\href{http://xxx.lanl.gov/abs/hep-ph/0204023}{{\tt
  hep-ph/0204023}}].

\bibitem{Aurenche:2006vj}
P.~Aurenche, M.~Fontannaz, J.-P. Guillet, E.~Pilon, and M.~Werlen, {\it {A New
  critical study of photon production in hadronic collisions}},  {\em
  Phys.~Rev.} {\bf D73} (2006) 094007,
  [\href{http://xxx.lanl.gov/abs/hep-ph/0602133}{{\tt hep-ph/0602133}}].

\bibitem{Bourhis:1997yu}
L.~Bourhis, M.~Fontannaz, and J.-P. Guillet, {\it {Quarks and gluon
  fragmentation functions into photons}},  {\em Eur.~Phys.~J.} {\bf C2} (1998)
  529--537, [\href{http://xxx.lanl.gov/abs/hep-ph/9704447}{{\tt
  hep-ph/9704447}}].

\bibitem{Chatrchyan:2012vq}
{\bf CMS} Collaboration, S.~Chatrchyan et~al., {\it {Measurement of isolated
  photon production in $pp$ and PbPb collisions at $\sqrt{s_{NN}}=2.76$ TeV}},
  {\em Phys.~Lett.} {\bf B710} (2012) 256--277,
  [\href{http://xxx.lanl.gov/abs/1201.3093}{{\tt arXiv:1201.3093}}].

\bibitem{Adare:2012yt}
{\bf PHENIX} Collaboration, A.~Adare et~al., {\it {Direct-Photon Production in
  $p+p$ Collisions at $\sqrt{s}=200$ GeV at Midrapidity}},  {\em Phys.~Rev.}
  {\bf D86} (2012) 072008, [\href{http://xxx.lanl.gov/abs/1205.5533}{{\tt
  arXiv:1205.5533}}].

\bibitem{Frixione:1998jh}
S.~Frixione, {\it {Isolated photons in perturbative QCD}},  {\em Phys.~Lett.}
  {\bf B429} (1998) 369--374,
  [\href{http://xxx.lanl.gov/abs/hep-ph/9801442}{{\tt hep-ph/9801442}}].

\bibitem{Miller:2007ri}
M.~L. Miller, K.~Reygers, S.~J. Sanders, and P.~Steinberg, {\it {Glauber
  modeling in high energy nuclear collisions}},  {\em
  Ann.~Rev.~Nucl.~Part.~Sci.} {\bf 57} (2007) 205--243,
  [\href{http://xxx.lanl.gov/abs/nucl-ex/0701025}{{\tt nucl-ex/0701025}}].

\bibitem{CMS:2013cka}
{\bf CMS} Collaboration, {\it {Charged particle nuclear modification factor and
  pseudorapidity asymmetry in pPb collisions at sqrt(sNN)=5.02 TeV with CMS}},
  {\em CMS-PAS-HIN-12-017} (2013).

\bibitem{Abelev:2013yxa}
{\bf ALICE} Collaboration, B.~B. Abelev et~al., {\it {$J/\psi$ production and
  nuclear effects in p-Pb collisions at $\sqrt{S_{NN}}$ = 5.02 TeV}},  {\em
  JHEP} {\bf 1402} (2014) 073, [\href{http://xxx.lanl.gov/abs/1308.6726}{{\tt
  arXiv:1308.6726}}].

\bibitem{Aad:2013lpa}
{\bf ATLAS} Collaboration, G.~Aad et~al., {\it {Measurement of the inclusive
  jet cross section in pp collisions at sqrt(s)=2.76 TeV and comparison to the
  inclusive jet cross section at sqrt(s)=7 TeV using the ATLAS detector}},
  {\em Eur.~Phys.~J.} {\bf C73} (2013) 2509,
  [\href{http://xxx.lanl.gov/abs/1304.4739}{{\tt arXiv:1304.4739}}].

\end{thebibliography}\endgroup

\end{document}